# Wirelessly transmitted subthalamic nucleus signals predict endogenous pain levels in Parkinson's disease patients

Abdi Reza[1,2], Takufumi Yanagisawa[1,3] *, Naoki Tani[1] **, Ryohei Fukuma[1,3], Takuto Emura[1], Satoru Oshino[1], Ben Seymour[4,5], Haruhiko Kishima[1]

[1]Department of Neurosurgery, Graduate School of Medicine, Osaka University, Osaka 565-0871, Japan.

[2]Department of Neurosurgery, Faculty of Medicine, Universitas Indonesia, Jakarta 10430, Indonesia.

[3]Institute for Advanced Co-Creation Studies, Osaka University, Suita, Osaka 565-0871, Japan.

[4]Institute of Biomedical Engineering, University of Oxford, Oxford OX37DQ, England, UK.

[5]Wellcome Centre for Integrative Neuroimaging, University of Oxford, Oxford OX39DU, England, UK.

**Corresponding author:** Takufumi Yanagisawa, Address: Department of Neurosurgery, Graduate School of Medicine, Osaka University, 2-2 Yamadaoka, Suita, Osaka 565-0871, Japan. Tel.: +81-6-6879-3652, E-mail: tyanagisawa@nsurg.med.osaka-u.ac.jp

**Abstract**

Parkinson's disease (PD) patients experience pain fluctuations that significantly reduce their quality of life. Despite the vast knowledge of the subthalamic nucleus's (STN's) role in PD, the STN biomarkers for pain fluctuations and the relationship between bilateral subthalamic nucleus (STN) activities and pain occurrence are still less understood. This observational study used data-driven methods by collecting annotated pain followed by a series of corresponding binary pain ratings and wirelessly transmitted STN signals, then leveraging the explainable machine learning algorithm to predict binary pain levels and sort the feature influence. The binary pain levels could be predicted among annotated pain reports corresponding to PD-related pain characteristics. The STN activity from both sides could impact pain prediction, with gamma and beta bands in the contralateral STN and delta and theta bands in the ipsilateral STN showing a prominent role. This study emphasizes the role of bilateral STN biomarkers on endogenous pain fluctuations.

## Introduction

Pain is a prominent nonmotor symptom of Parkinson's disease (PD)[1] that results from complex pathological processes involving the central and peripheral nervous systems.[1–4] Pain in PD patients can be categorized as PD-related pain (PDRP) or non-PDRP.[1,2,4,5] Patients may also report pain in multiple locations,[2,5] leading to multiple "annotated pain" reports that make each individual's endogenous pain experiences unique.[2] The prevalence of pain among PD patients ranges from 40% to 80%,[1,2,6] yet it remains underrecognized and undertreated.[1,5,6]

Like motor symptoms, pain in PD patients can fluctuate throughout the day.[6–10] These fluctuations are associated with ON-and-OFF levodopa periods, motor symptom variability, younger age at disease onset, and greater disease severity.[6–8,10] Pain fluctuations negatively impact quality of life (QoL),[10,11] but their identification[7] and severity ratings[1,9] still rely primarily on self-reports. Recognizing the severity levels of pain fluctuation through objective biomarkers is imperative for pain management.

Previous studies have suggested that local field potentials (LFP) from the STN could provide insights into pain occurrence. For example, dopaminergic activities have been shown to modulate pain in both clinical and laboratory studies.[12,13] It shows that beta,[14,15] and gamma[16] bands are identified as the most affected by dopamine fluctuations, and a machine learning (ML) classifier supplied with STN's LFP features could predict the dopaminergic states.[17] The influence of motor symptoms on pain fluctuations has also been well-documented.[1,3,6–8,10] STN-LFP also represents the severity of overall motor symptoms, especially in gamma and low beta bands.[16,18] These findings highlight the potential role of STN-LFP as an objective biomarker of endogenous pain fluctuations. Furthermore, experimental studies have demonstrated that exogenous pain induces alpha,[19] or beta activity[20] in the STN. Anatomically, the subthalamic nucleus (STN) has been linked to pain processing and essential connections to the brainstem, subcortical, and cortical regions involved with endogenous pain modulation.[3] However, whether endogenous pain fluctuations are affected by a single or both STNs is still unclear; thus, further exploration is needed to unveil the bilateral STNs' roles in pain occurrence.

Interestingly, STN-targeted deep brain stimulation (DBS), a surgical intervention for PD motor symptoms, has also been shown to reduce pain symptoms.[9,21] Modern adaptive DBS systems, which deliver intermittent stimulation based on detected biomarkers, offer similar efficacy to continuous DBS for PD treatment but with fewer side effects and lower battery consumption.[22] Nowadays, adaptive DBS systems can transmit STN signals wirelessly.[23] Thereby, infections associated with externalized wires[24] are avoidable, and an extended measurement period becomes common practice.[17,25] However, compared to PD motor fluctuations, adaptive DBSs' role in treating pain fluctuations remains limited due to the poor understanding of STN biomarkers for pain; identifying pain fluctuation biomarkers from both STNs could expand the current application of adaptive DBSs.

In this study, we hypothesized that the binary levels representation of pain fluctuations ratings could be predicted from bilateral STN-LFP activity using a machine learning (ML) classifier for each annotated pain reported by patients. Segmenting widely spread pain ratings into binary[26] or lesser levels[27] is a practice to reduce the complexity of the decoding task[27] and address the non-linearity of pain perception.[26] We consider the bilateral STN features necessary for comprehensive model interpretation of whether pain perception is influenced by a single or both STN. The ML model training for individualized reports rather than pooled reports has been adopted and widely used in endogenous pain research[26] and PD research about dopamine[17] and sleep[28] states. This individualized ML approach takes benefits from report-specific variability and supports personalized medicine.[17,26] Furthermore, given that pain is associated with patients' motor symptoms, mood, and fatigue,[10,11,13] we aimed to evaluate model performance via these alternative features accompanying the STN-LFP features. To achieve this goal, we collected annotated pain reports from each patient before surgery. After DBS system implantation, this annotated pain report was expanded into a series of repeated observations by collecting the corresponding pain ratings and other self-ratings (consisting of motor symptoms, mood, and fatigue questions) and wirelessly transmitted STN-LFP signals during the same visit. A random forests (RF) was then used as an explainable ML algorithm to predict binary pain levels of each annotated pain report and identify influential frequency components in each STN,[29,30] potentially uncovering the influential electrophysiological biomarkers and their corresponding STN sides.

**Results**

Eleven patients who underwent two steps of unilateral or one step of bilateral DBS electrode implantations agreed to participate in the study. Thirty-one pain-annotation reports with a series of corresponding pain VAS ratings, STN-LFP features, motor symptoms features, mood, and fatigue features were obtained from patients; the 0-100 VAS ratings were then transformed into binary classes using each series' median value. Twenty-three pain-annotation reports must be excluded from ML training either due to imbalanced class after median splitting (n = 16) or only unilateral electrodes implanted at the time of repeated observations (n = 7) (see Fig.1 for Strengthening the Reporting of Observational Studies in Epidemiology (STROBE) flowchart). The omission of collected reports in observational studies dealing with endogenous pain fluctuations is sometimes unavidable. In contrast to the experimental study, we had no control over patients' perceptions, ratings, or distribution of pain experience at each level.[27] Especially in the absence of criteria for pain fluctuation, researchers define their criteria and omit the unfit observations.[31]

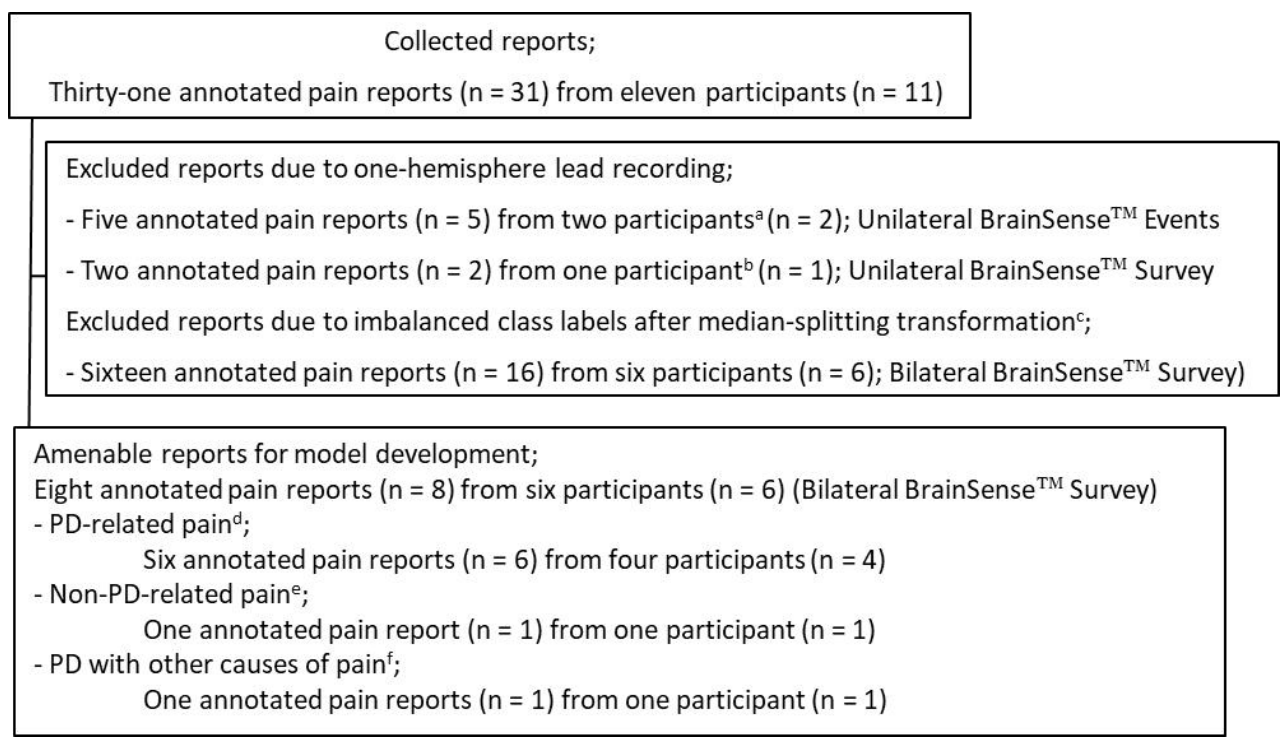

**Figure 1.** STROBE flow chart (STROBE: Strengthening the Reporting of Observational Studies in Epidemiology).

a, b: Participants who initially contributed unilateral recordings were readmitted for second implantation surgery and reinvited for signal recordings with bilateral LFP recording using BrainSense™ Survey.

c: Annotated pain reports with imbalanced labels are not amenable for model training and therefore excluded.

d: PD-related pain was determined from patient history and physical examination, based on key information related to pain occurrence related to PD symptoms, influence of PD symptoms, and influence of PD medications.

e: One patient experiences pain following a femoral-hip injury followed by surgical treatment.

f: One participant had no consistent pain complaints before the DBS implantation. The participant was asked to report any pain from the body in general

---

In PD, pain fluctuations were determined mainly from the patient's perception,[7,9] which is often obscure. In this study, we transformed the observational series of VAS ratings into binary levels (higher and lower pain levels) using its median values. We define fluctuated pain reports as those with a relatively balanced distribution to both levels following median splitting; these annotated pain series were retained for ML model training. In contrast, the imbalanced annotated pain series in which any binary level occupied more than 60% of the observations[32] were considered as 'lack of fluctuation' or 'nonfluctuation' pain reports; they did not need a prediction model and were therefore excluded (see Supplementary Fig. S1). Nevertheless, all reports are still helpful for clinical evaluation and constructing a patient's follow-up library as part of a larger PD study in our center.

The annotated pain reports for model training consist of eight reports (n = 8) from six patients (n = 6) (four females; age 69.00 ± 3.03 years [mean ± standard deviation (SD)]; PD symptom duration 14.00 ± 4.24 years [mean ± (SD)]). We evaluate whether the annotated pain reports correspond to PDRP or non-PDRP characteristics (see Fig.2 and Supplementary Table S1), thus associating six annotated pain reports with PDRP (n = 6) and the other two with non-PDRP (n = 2). We also evaluate the pain area characteristics and their laterality relative to the STN location. The overall demographic of patients and annotated pain reports are described in Demographic Table 1.

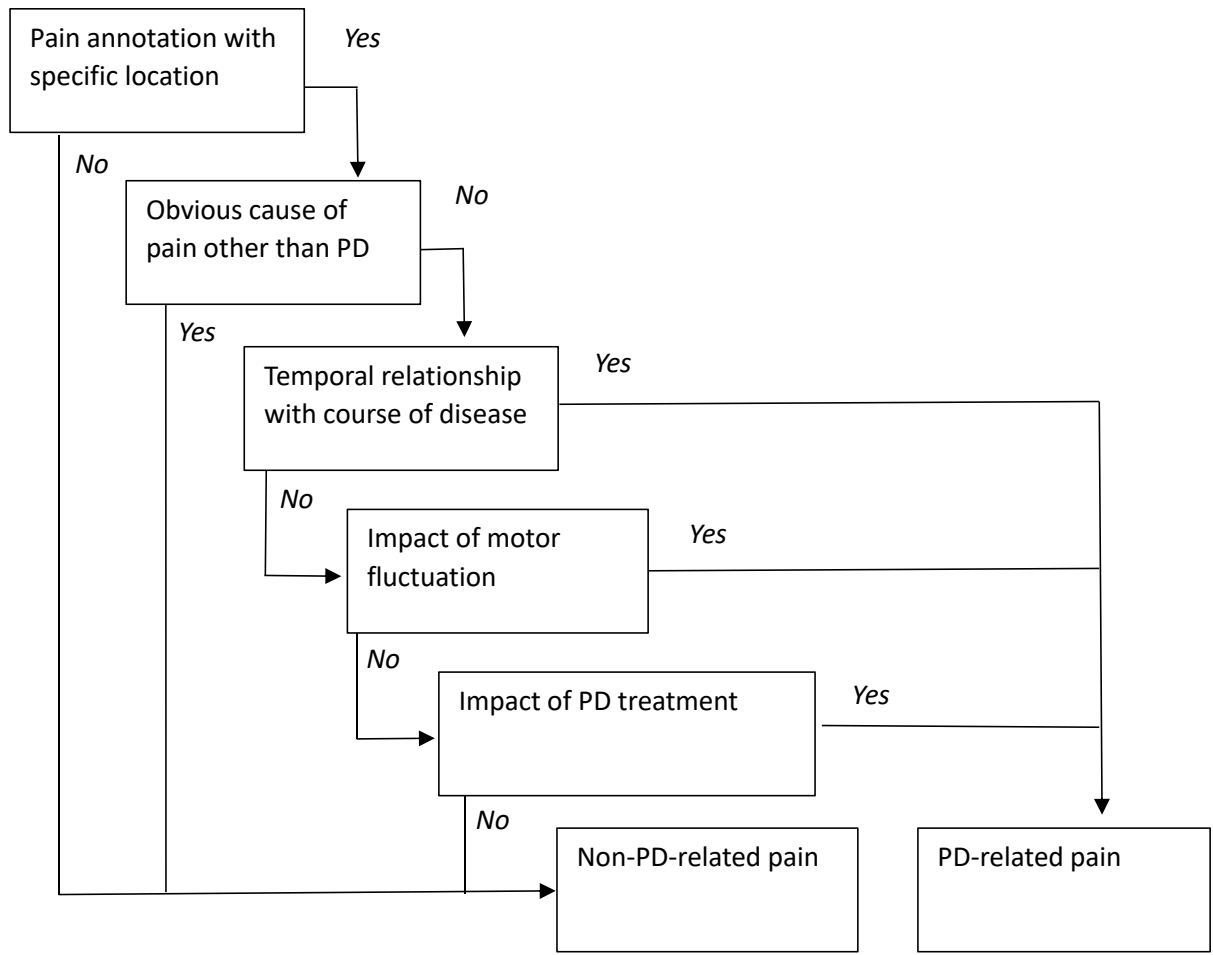

**Figure 2.** Modified criteria for evaluating Parkinson's disease-related pain (PDRP) association types. Pain without a specific location was separated from PDRP since it could occur due to any cause. Pain arising from other identifiable causes was separated from PDRP. Pain was categorized as PDRP if it had a temporal association with PD, was impacted by PD motor fluctuations, or was influenced by PD treatment effects.

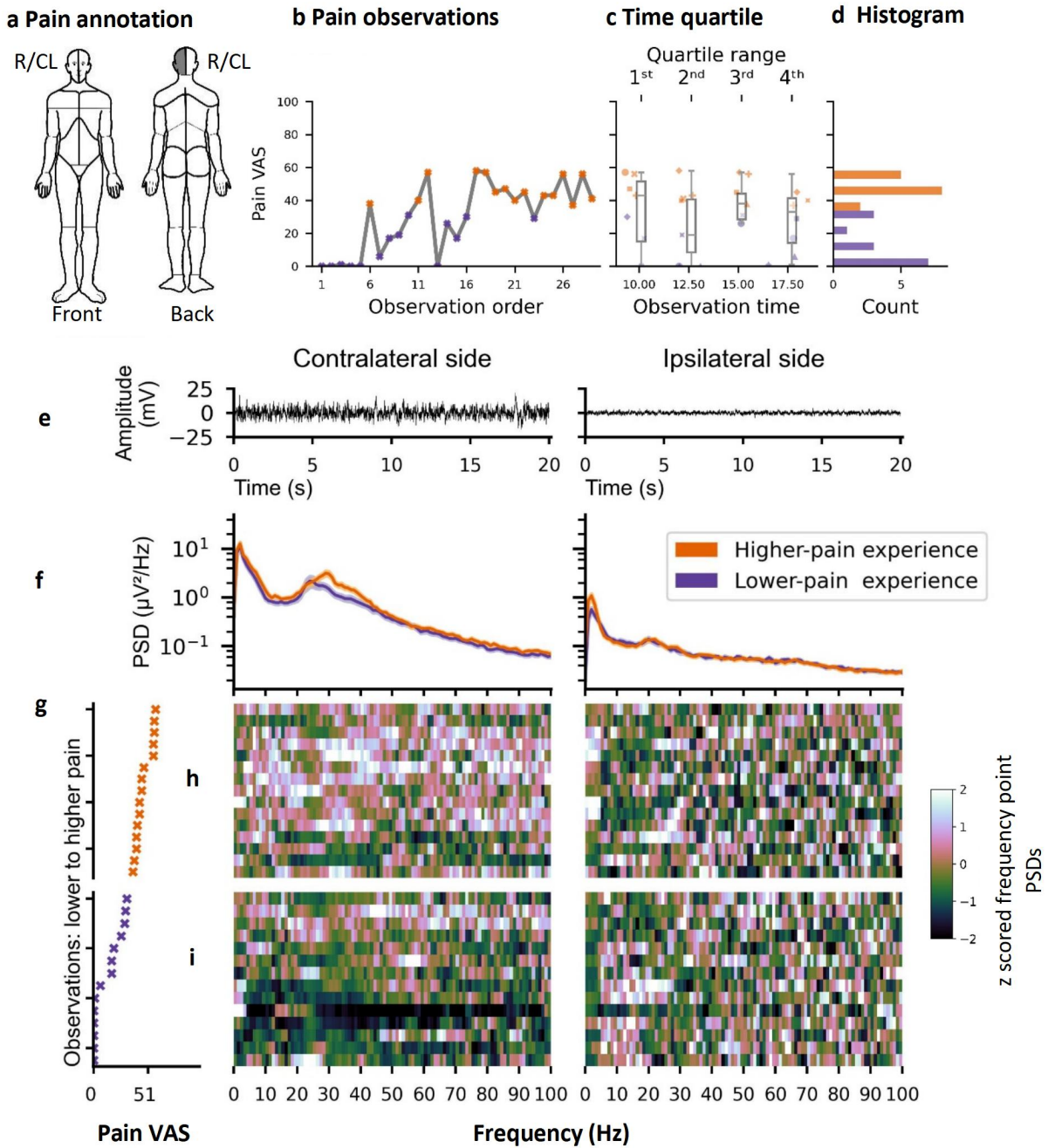

**Figure 3.** Overview of pain fluctuation report with lateral and specific location and PDRP characteristics (annotated pain report #E). **a** Pain map showing self-reported pain location (left occipital), with body pain areas referenced from Margolis et al.[33] Grey shading indicates pain areas (R: right; CL: contralateral side of pain annotation). **b** Pain fluctuations shown by observation order. **c** Scatterplot of pain ratings by daily recording time; same-day recordings are represented by consistent symbols. Data points are overlaid with boxplots indicating central tendency and distribution across four quartiles of the time distribution: median (centre bar), first and third quartiles (lower and upper box borders), and whiskers extending to 1.5 times the interquartile range. No significant difference was found across the four-quartile pain distributions (annotated pain report #E; $F(28) = 0.29$, $P = 0.83$; one-way ANOVA). **d** Histogram showing the distribution of pain ratings, categorized into higher and lower pain classes. **e** STN time series signals correspond to recording site. **f** Average power spectral density (PSD) amplitudes of higher and lower pain experiences on a log scale (y-axis); the shaded area represents the standard error of the mean (SEM). **g** Observations reordered from lower to higher pain. (H, I) Heatmaps of z score PSDs for reordered observations, showing frequency-specific activity in higher (H) and lower (I) pain states

**Table 1.** Demographics and characteristics of the included patients.

| | Annotated pain reports | | | | | | | |
|---|---|---|---|---|---|---|---|---|
| | **A** | **B** | **C** | **D** | **E** | **F** | **G** | **H** |
| Patient | P1 | | P2 | P3 | P4 | | P5 | P6 |
| Age at data collection | 70's | | 60's | 70's | 70's | | 70's | 70's |
| SFMPQ2 (total) | 0.29 | | 0.03 | 0.19 | 0.2 | | 0.05 | 0.09 |
| PFS (Fatigue/not fatigue) | 3.13 (Not Fatigue) | | 4.13 (Fatigue) | 3.75 (Fatigue) | 3.88 (Fatigue) | | 4.38 (Fatigue) | 3.31 (Fatigue) |
| Neurostimulator model | Percept PC B35200 | | Percept PC B35200 | Percept PC B35200 | Percept PC B35200 | | Percept PC B35200 | Percept PC B35200 |
| Lead model | 3389 | | 3389 | B33005 | B33005 | | B33005 | 3389 |
| Surgical stages | Two stages | | One stage | One stage | One stage | | One stage | Two stages |
| Pain annotation | Left back | Right back | Back of the neck | Left lower back | Left occipital | Back of the neck | Body in general | Left abdomen |
| Location categorization | Lateral and specific | Lateral and specific | Non-lateral and specific | Lateral and specific | Lateral and specific | Non-lateral and specific | Nonspecific | Lateral and specific |
| PD-pain characteristics | PDRP | PDRP | PDRP | PDRP | PDRP | PDRP | Non-PDRP | Non-PDRP |
| VAS lower class; Number of trials (VAS range) | 24 (13-24) | 22 (0-22) | 17 (0-0) | 11 (13-29) | 14 (0-31) | 13 (0-30) | 20 (19-46) | 35 (0-32) |
| VAS higher class; Number of trials (VAS range) | 18 (25-85) | 20 (23-75) | 13 (1-55) | 10 (31-73) | 15 (37-58) | 16 (34-60) | 17 (48-83) | 33 (33-88) |
| Amount of observations | 42 | 42 | 30 | 21 | 29 | 29 | 37 | 68 |
| Maximum observations per day | 4 | 4 | 4 | 4 | 4 | 4 | 4 | 4 |
| Average (SD) observation per day | 3.82 (0.40) | 3.82 (0.40) | 3.75(0.46) | 3 (0.58) | 3.63 (0.74) | 3.63 (0.74) | 3.7 (0.67) | 4 (0) |
| Average (SD) interval between observations per day-in hours | 3.06 (0.66) | 3.06 (0.66) | 2.80 (0.55) | 3.28 (0.79) | 2.78 (0.43) | 2.78 (0.43) | 2.69 (0.77) | 2.57 (0.79) |

Abbreviations: PFS, Parkinson's Fatigue Scale; PD, Parkinson's disease; PDRP, Parkinson's disease-related pain; SD, standard deviation; SFMPQ2, Short-Form McGill Pain Questionnaire-2; VAS, visual analogue scale. Pain and patients are described for presentation purposes; neither the alphabetical nor the numerical representation of the patient and report are associated with the patient's initial or the recruitment order.

**Report-specific patterns are derived from spectral characteristics of each pain level**

Before model training, each annotated pain report was inspected separately. For example, a representative pain report #E, showed pain in the left occipital area (Fig. 3a). A series of pain ratings are displayed in sequential order (Fig. 3b) or descending order (Fig. 3g). Scatterplot shows the pattern of VAS ratings and time stamp of report collection in the same day, accompanied with boxplot representing VAS distribution into four groups of time stamping; four is the maximum repeated session in a day (Fig. 3c), there is no significant difference across four quartile groups ($F(28) = 0.29$, $P = 0.83$; one-way ANOVA). A horizontal histogram displayed the distribution of higher and lower pain levels after median splitting (Table 1, Fig. 3d), showing a balanced distribution of binary pain levels. The time series signals from three pairs of electrodes (0-1, 1-2, 2-3) of each side, collected by BrainSense™ Survey via the Medtronic clinician programmer tablet[23] were then reduced into single time series by retaining the first principal component (PC1) following principal component analysis (PCA) (Fig. 3e). These PC1 time series are converted into power spectral density amplitudes using Welch's methods (Fig. 3f), different peaks amplitude average was shown for higher and lower pain levels. Contralateral electrodes show higher pain levels corresponding to an average of the higher delta, theta, alpha, low beta, high beta, low gamma, and high gamma amplitudes. Ipsilateral electrodes show higher pain levels corresponding to an average of the higher delta amplitudes. The pain ratings are then displayed in descending order (Fig. 3g), accompanied by the corresponding z-scored PSD peak amplitudes within the same frequency points inside a heatmap of higher (Fig. 3h) and lower pain (Fig. 3i) levels.

Heatmaps describe the association of pain levels and z-scores of peak amplitudes from each observation. See Supplementary Fig. S2-S8 for an inspection of other annotated pain reports.

For each annotated pain report and each side, the average PSD amplitude peaks in the following frequency bands: delta (1–4 Hz), theta (5–8 Hz), alpha (9–12 Hz), low beta (13–20 Hz), high beta (21–30 Hz), low gamma (31–60 Hz), and high gamma (61–90 Hz), are prepared as features for model training. A two-tailed Welch's test was implemented to compare the difference of the canonical band PSD's average distribution between higher and lower pain levels. It shows that each report has specific discriminative STN-LFP patterns for higher and lower pain levels. For example, pain report #E has PDRP characteristics and shows a higher pain level corresponding significantly ($P$value, Welch's test) to a higher PSD average of the ipsilateral theta (0.002), contralateral theta (0.005), alpha (0.001), low-beta (0.007), low-gamma (0.000), and high-gamma (0.008) with significant difference ($P$< 0.05, Welch's test). In contrast, pain report #H has non-PDRP characteristics and shows a higher pain level corresponding to a higher PSD average of the ipsilateral delta (0.377), theta (0.709), alpha (0.975), and contralateral high-beta (0.693), low-gamma (0.376), and high gamma (0.687) without significant difference ($P$> 0.05, Welch's test). The functionality of these report-specific discriminative STN-LFP patterns is then challenged for binary pain decoding tasks. For visualization purposes, this canonical band average was converted into a z-score, and the comparison of distribution between canonical bands into higher and lower pain levels was displayed with paired boxplots (Fig. 4).

In addition, we also collected ratings for five questions about mood and fatigue and five questions about motor symptoms to construct alternative features (see Supplementary Table S2 for these alternative features and Supplementary Fig. S9 for the distribution of motor symptoms, mood, and fatigue rating into binary pain levels).

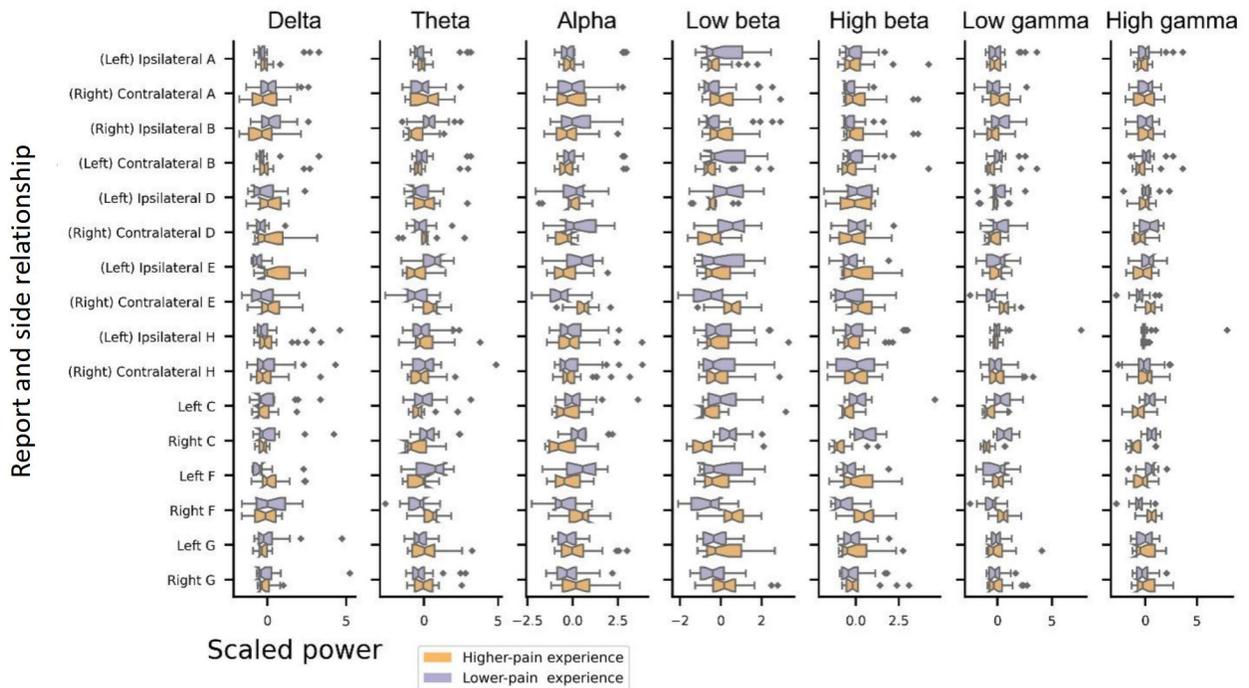

**Figure 4.** Boxplots comparing the z-scored average canonical band PSD to the corresponding pain severity class labels. The boxplots illustrate the distribution of each z-scored PSD, with the median indicated by the middle bar within the box. The lower and upper box borders represent the first and third quartiles, respectively. Whiskers extend to 1.5 times the interquartile range, with separate diamonds (◊) indicating outlier data points outside the whiskers. The plots display annotated pain reports' variability, comparing the canonical band's distribution across higher (warmer colors) vs. lower (colder colors) pain levels.

**STN-LFP features effectively classified higher or lower pain experiences with PDRP characteristics**

To obtain model prediction, Random forests (RF) with nested cross-validation (CV) were trained for each annotated pain report using the bilateral non-normalized values of averaged PSD peak amplitude within the aforementioned canonical bands. The non-normalized average of PSD peaks within the canonical band was chosen since normalization could introduce a particular influence of 'information leakage,' shift the test set distribution, and add obstacles for real-time prediction.[34] We chose the random forest classifier as an algorithm that does not need feature normalization and can sort feature influence.[29,30]

Among the six annotated pain reports with PDRP characteristics, LFP features yielded significant decoding performance (Fig. 5, Supplementary Fig. S10) consisting of (average of balanced accuracy (*P*value, permutation test)): 68.625%(0.018), 86.25%(0.001), 74.583%(0.012), 71.667%(0.016) for four annotated pain reports #B, #C, #E, and #F, respectively. The non-significant decoding performance consists of 55.125%(0.253) and 51.667%(0.401), despite an above 50% balanced accuracy performance, obtained from two annotated pain reports, #A and #D, respectively. Nevertheless, group-level analysis of those six annotated pain with PDRP characteristics revealed a mean balanced accuracy of 67.99 ± 12.83% (mean ± SD), which was more significant than the chance level ($P = 0.0156$, n = 6, one-sided Wilcoxon signed-rank test, Bonferroni-adjusted α = 0.0167 [0.05/3]; Fig. 5).

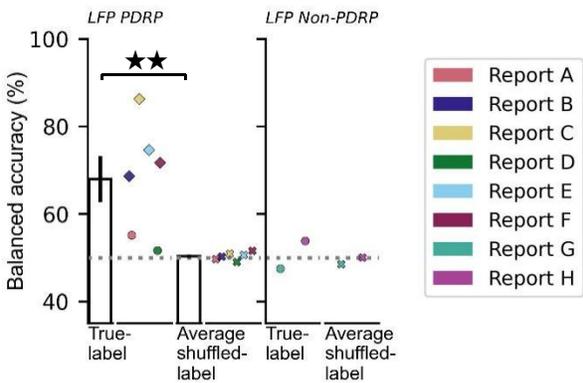

**Figure 5.** Average balanced accuracy across predictor subsets. The balanced accuracy values for each annotated pain report are shown in distinct colors. A diamond (◊) indicates annotated pain reports with above-chance performance from a permutation test, a circle (○) indicates annotated pain reports not surpassing the permutation test threshold, and a cross (x) shows the average performance of 1000 label-shuffled iterations. The bars represent group-level averages for the PDRP annotated pain reports, with error bars indicating the standard error of the mean (SEM). A double asterisk (★★) denotes the results of a one-sided Wilcoxon signed-rank test against each annotated pain report's corresponding null distribution average, with a Bonferroni-corrected significance threshold of $P < 0.0167$ (0.05/3). A single asterisk (★) indicates $P < 0.05$.

In each of the two annotated pain reports with non-PDRP characteristics, the STN-LFP features show non-significant results of the average balanced accuracy ($P$ value, permutation test): 47.5%(0.550) and 53.81%(0.261) for reports # A and #D, respectively.

To verify our STN-LFP model, additional analyses of intraindividual and group-level model performance derived from alternative features, including motor symptoms, mood, and fatigue features, are presented in Supplementary Fig. S10, S11.

**Bilateral STN contributes to pain prediction.**

To evaluate the influence of canonical bands from both STN, we utilized Gini importance, a common measurement of how influential a feature is for the individual decision tree (DT) split inside an RF model. Among the annotated pain reports with significant STN LFP model performance, the contralateral STN presented greater Gini importance based on the average across frequency band features than did the ipsilateral STN for annotated pain reports distributed on either side of the body (lateral and specific annotated pain) (report #E and #B; Fig. 6a). For pain annotations centered in the body (non-lateral and specific annotated pain), the right STN demonstrated more prominent Gini importance on average across frequency band features (report #C and #F; Fig. 6b). In annotated pain reports with higher average Gini importance, the beta and gamma bands had a prominent influence (report #B, #C, #E, and #F; Figure 6b). Conversely, for the STN with lower average Gini importance, the low-frequency bands (delta and theta) had a prominent influence (report #B, #E, and #F; Fig. 6c). See also Supplementary Fig. S12 for the electrode-specific model performances. For group level Gini importance of motor symptoms, mood, and fatigue features, please see Supplementary Fig. S13 and Supplementary Fig. S14– S21 for each annotated pain report's feature influence based on Gini importance and SHAP.

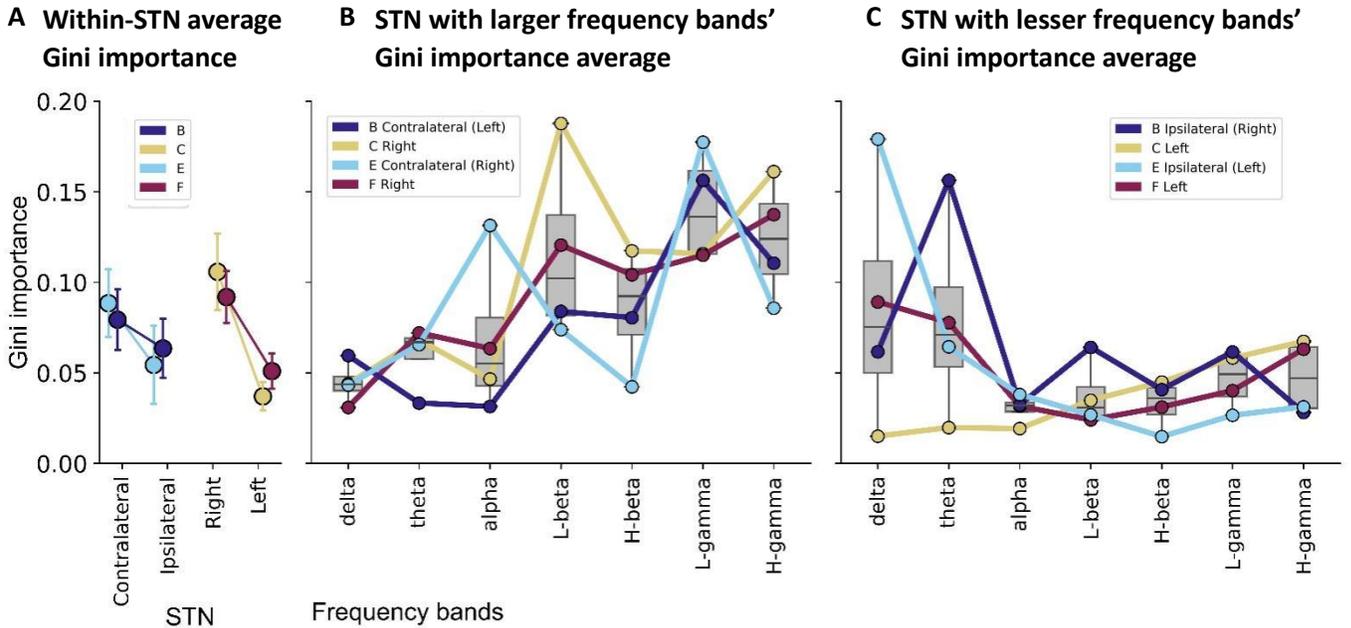

**Figure 6.** Group-level STN-LFP Gini importance of annotated pain reports. Annotated pain reports #B, #C, #E, and #F represent annotated pain reports with significantly balanced accuracy performance using STN-LFP predictors. **a** The Gini importance averages show that the contralateral STN in annotated pain reports #B and #E has greater Gini importance than the ipsilateral STN. Similarly, the right STN in annotated pain reports #C and #F has greater Gini importance than the left STN. The circular points represent the average Gini importance on the corresponding x-axis, with error bars indicating the standard error of the mean (SEM). **b** Distribution of Gini importance for frequency bands in the contralateral STN for annotated pain reports #B and #E and the right STN for annotated pain reports #C and #F. **c** Distribution of Gini importance in the ipsilateral STN for annotated pain reports #B and #E and the left STN for annotated pain reports #C and #F. Box plots show the distribution of Gini importance values, with the median represented by the middle line, first and third quartiles as the lower and upper box borders, and whiskers extending to 1.5 times the interquartile range.

**Discussion**

In this study, we identified three main findings regarding the role of STN-LFP in predicting pain fluctuations among patients with PD. First, STN-LFP features can predict pain levels from annotated pain associated with Parkinson's disease-related pain (PDRP). Second, the model features derived from STN-LFP were comparable to the alternate ML model, which incorporates motor symptoms and is further complemented by mood and fatigue features (see online supplementary discussion). Third, the bilateral STN-LFP exhibited an influence on pain prediction. To our knowledge, this is the first study that attempted to decode endogenous pain fluctuation and explore the different roles that bilateral STN has. These findings underscore the significance of STN activities and warrant further exploration in the context of endogenous pain fluctuations.

Interestingly, STN-LFP features predict pain annotations corresponding to the PDRP group, suggesting that the apparent relationship between pain and PD is not only crucial for therapeutic response,[35] yet also for our proposed pain fluctuation model. Post implantations, several studies have shown that pain relief resulting from DBS does not always align with improvements in motor symptoms.[36,37] This discrepancy may be attributed to the DBS effect on the central modulation of pain perception,[37] as somatosensory processing enhances[38] and increases the detection threshold of thermal and mechanical stimuli.[36] These observations suggest diverse, albeit potentially overlapping, mechanisms underlying motor symptoms and pain, especially in PDRP patients.

Our modified PDRP group (see Fig. 2 and Table S1 for our modified criteria for PD-associated pain) consisted of patients with apparent pain associated with PD and showed specific pain locations.[1,2,4,5] While our explainable machine learning analysis revealed that frequency bands strongly influence pain prediction, these associations do not imply causality. The identified frequency bands may directly represent pain modulation activity, indirectly reflect pathological motor activity, or signify shared underlying processes such as dopamine depletion. Further research is needed to elucidate the specific relationships between STN frequency bands and motor and pain fluctuations in PD patients.

Additionally, while our several models showed significant predictive ability, some model underperformance suggests that other factors not captured by STN-LFP alone may contribute to pain perception in PD patients (see online supplementary discussion for motor symptoms, mood, and fatigue model performance). Among eight collected annotated pain reports (n=8), two reports (n=2) cannot be predicted either by STN-LFP, motor symptoms, or mood and fatigue features (report #D, #G, Supplementary Fig. 10). Considering unexplained pain was found among 11.6% of adults with chronic pain,[39] further attention is necessary to determine if this non-decodable pain is associated with factors outside our incorporated features or is due to subjective interpretation of pain. Two other annotated pain reports (n=2) could be decoded with mood and fatigue features (report #A, #H, Supplementary Fig. 10); these represent pain fluctuation reports that are more associated with mood and fatigue than STN activities. The rest of the annotated pain reports (n=4) that have PDRP characteristics could be decoded with above-chance level performance ($P$value < 0.05, permutation test). We consider these #B, #C, #E, and #F reports to be invaluable in determining a meaningful interpretation of the canonical band influence and its corresponding STN to the fluctuations of pain.

To evaluate the bilateral STNs' role in pain occurrence, we look at the Gini importance of each STN. For lateral and specific annotated pain, the contralateral STN demonstrated greater importance in the model (Fig. 6a), especially in the beta and gamma frequency bands (Fig. 6b). This observation aligns with the somatotopic organization of the sensorimotor system,[40] the association between contralateral STN signal activity and the site of experimentally induced pain,[20] and the association between contralateral STN activity and the most affected body symptoms.[41] Furthermore, the right STN demonstrated greater importance for centrally distributed pain (non-lateral and specific annotated pain), suggesting hemispheric specialization in pain processing,[42] which warrants further investigation.

Interestingly, our analysis indicated that the ipsilateral STN also contributes to pain prediction, primarily through the delta and theta frequency bands (Fig. 6c). The connectivity between the bilateral STNs is supported by mechanisms such as the STN-cortex-STN loop,[43] or throughout the more extensive system connecting limbic, associative, and motor networks.[44] These mechanisms may contribute to the role of the bilateral STN in pain fluctuations. Consequently, we propose that the contribution of bilateral STN activity should not be overlooked in the context of pain modulation. This bilateral involvement suggests a complex interplay between STN activity and pain perception, potentially reflecting the widespread nature of pain processing in the brain,[40] or potentially a more direct link between STN and descending pain modulatory centers.[3,13] Future studies should investigate the specific roles of the contralateral[20,41] and ipsilateral STN in pain modulation and whether this bilateral involvement could be leveraged for more effective pain management strategies in PD.

This study highlights the role of STN-LFP as the biomarker for pain fluctuation and corresponding bilateral STN contribution. Based on these findings, a pain-specific ML model can alert clinicians to the severity of pain experienced by PDRP patients, facilitating timely intervention. The significant STN-LFP model prediction also objectively confirms the association between pain occurrence and PDRP characteristics. The explainable ML model, which leverages Gini importance (see Fig. 6, Supplementary Fig. S14-S22 for SHAP values), provides insights into the most influential features and their directional impact. For example, in annotated pain report #E, contralateral low-gamma and ipsilateral delta activity emerged as the most influential features, indicating its promise as an individualized biomarker for pain fluctuations and a personalized target for closed-loop DBS systems.

Despite the invaluable insights this research provides, our study has certain limitations. Since the annotated pain series were collected shortly after bilateral implantation during hospitalization, peri-lead edema may have influenced the signal recording.[45] Therefore, it would be valuable to conduct a future longitudinal study to assess the stability of these biomarkers over time and their associations with the progression of pain and motor symptoms. As a single-center study with a relatively small sample size, our findings should be interpreted cautiously and may not be generalizable broadly. The limited number of non-PDRP patients was insufficient to draw assertive conclusions, highlighting the need for larger-scale validation studies. Furthermore, STN-LFP data collection was restricted by the 250 Hz sampling rate of the BrainSense™ Survey,[23] potentially constraining the exploration of other PD biomarkers, such as high-frequency oscillation (HFO) power and phase-amplitude coupling of beta and HFO.[41] The BrainSense™ Survey signal was collected with OFF-stimulation mode. Thus, the replicability of this finding with the ON-stimulation condition needs further exploration. However, the BrainSense™ Survey, as we implemented in this study, requires only minimal battery consumption while allowing real-time signal transmission across canonical frequency bands.

Individualized medication, stimulation, or rehabilitation therapies were administered as part of hospital care; as an observational study, we did not interfere with responsible clinicians' decisions. This may increase patient variability and certain confounding factors not included in the feature set. It also limits natural pain cycles, meaning only a subset of the annotated pain reflected distinct fluctuations over repeated visits. However, strict control, including prolonged OFF medication, may lead to patients' burdens increased, satisfaction reduced, and compliance descended. Patient improvement, satisfaction, and compliance are still the utmost priority in our observational study. Reports-specific ML model training was chosen instead of between-subject model implementation to deal with interindividual variability confounders, as commonly utilized in brain-computer interface (BCI) research. To avoid the 'curse of dimensionality,' we incorporated only a sensible number of features to train the model. Nonetheless, our study offers the ML model for pain fluctuations under conditions that closely resemble real-life scenarios.

In conclusions, this study finds that the LFP of the STN stores information that represents the binary class of pain fluctuation severity among PD-related pain patients. The frequency band PSD in the STN is suggested to be a biomarker for PD-related pain, and the bilateral STN plays a role in pain modulation.

**Methods**

This study was conducted at Osaka University Hospital from October 2020 to March 2022. A convenience sample was obtained from eligible patients scheduled for surgical implantation of an adaptive DBS system with wireless data transfer, the Percept™ PC platform (Medtronic, Minneapolis, MN, USA). Recruitment was based on recommendations from the surgeons overseeing patient care. The study received approval from the local institutional review board (approval no. 20213) and adhered to the Declaration of Helsinki. Written informed consent was obtained from all patients after they received explanations of the nature and potential consequences of the study.

This prospective observational exploratory study was conducted as part of a broader adaptive DBS study. This study did not interfere with the responsible clinicians' decisions concerning hardware selection, stimulation parameters, daily medication, or hospitalization duration (see Fig. 7 and Supplementary Fig. S22 for details on the data collection process). This study adhered to the Strengthening the Reporting of Observational Studies in Epidemiology (STROBE) guidelines.[46]

**Inclusion and exclusion criteria**

Participants with Parkinson's disease (PD) who were candidates for deep brain stimulation (DBS) therapy were recruited for this study. Patients received advice regarding DBS implantation based on medical staff consensus independent of the study. The inclusion criteria were as follows: male or female, aged 20–75 years, diagnosed with PD, deemed suitable for DBS implantation as recommended in a clinical meeting, and compatible with the Percept™ PC device (Medtronic, Inc.) for implantation.

Additional requirements included an American Society of Anaesthesiologists (ASA) preoperative status of 1, 2, or 3, good general health, adequate oral and written communication skills, the capacity to provide informed consent, and willingness to participate.

The exclusion criteria consisted of a diagnosis of movement disorders other than PD (e.g., essential tremor), an ASA status of 4 or 5, pregnancy, a history of incompatible implants that preclude MRI evaluation, and allergies to materials used in DBS implantation. Participants could also be excluded based on the physician's assessment if medical or nonmedical issues were identified (e.g., commitment, cooperativeness, communication skills, familiarity with digital devices, or increased risk/lack of benefit from repeated evaluations). No exclusions were based on race, sex, religion, or nationality. The participants retained the right to withdraw from the study at any time.

**Pain annotation and categorization**

Participants were assisted in mapping pain locations that continuously affected them prior to surgery with a body surface system.[33] They were allowed to freely indicate areas without being limited by the outlined figure (e.g., lower-left abdomen). Participants reporting multiple pain locations were asked to document each area. Those without specific pain locations were still included by documenting pain scores from their body in general (i.e., any pain experienced anywhere). Pain annotations were categorized as lateral and specific locations (e.g., left occipital, left abdomen), nonlateral and specific (e.g., back of the neck), or nonspecific locations (e.g., anywhere on the body). The annotated pain reports were evaluated for PDRP and non-PDRP characteristics, following our modified criteria from previous studies.[1,2,4,5] (see Fig. 2 and Supplementary

Table S1 for our modified criteria for PD-associated pain). Additional patient characteristics, including age, sex, Short Form of McGill Pain Questionnaire 2 (SFMPQ2) score,[47] and Parkinson's Fatigue Scale (PFS) score,[48] were recorded.

**Surgical procedure and postsurgical data collection**

The DBS system was implanted following the surgical procedures at Osaka University Hospital, as previously described.[49] In brief, the trajectory for cortical puncture and STN access was planned via preoperative magnetic resonance (MR) images fused with stereotactic frame-attached computed tomography (CT) scans via StealthStation (Medtronic, Minnesota, MN, USA) to reach the dorsolateral aspect of the STN. STN locations were confirmed via microelectrode recording and intraoperative X-ray visualization. Patients were implanted with a four-lead DBS electrode compatible with the Percept™ PC device (Medtronic, Minnesota, MN, USA). The decision to perform either two separate surgical sessions for unilateral electrode implantation or one session for bilateral electrode implantation and the choice between local and general anaesthesia were made on a case-by-case basis.

Data collection commenced once patients returned to the common ward and received clearance from the overseeing surgeon. This usually occurs one day after surgery but could be later based on the patient's wellness. Data collection visits were conducted a maximum of four times per day, interspersed with rest periods, other procedures, and ongoing hospital treatments. Data collection included recording the patient's corresponding VAS ratings of annotated pain, self-rating motor symptoms, mood, and fatigue, and STN LFP signal recordings.

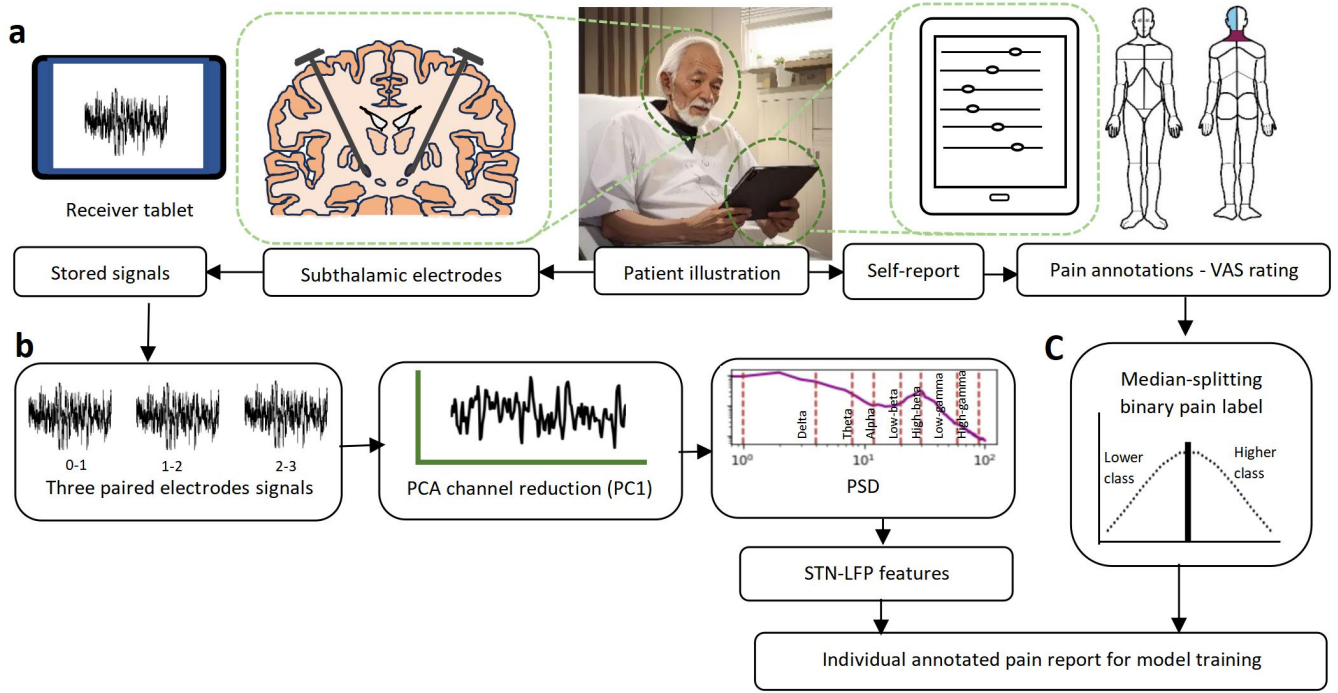

**Figure 7.** Schematic representation of intraindividual multimodal annotated pain report collection for binary pain fluctuation classification. **a** Annotated pain ratings were collected via a tablet interface (note: patient illustration was derived from a text-to-image artificial intelligence tool via www.fotor.com). Subthalamic nucleus (STN) electrodes, implanted in the patient, wirelessly transmitted signals via the Percept™ PC deep brain stimulation (DBS) system (Medtronic, Inc.), with activation in BrainSense™ Survey mode. **b** Signals from each STN hemisphere, comprising three electrode pairs (0-1, 1-2, and 2-3), were captured via the BrainSense™ Survey. A single time series was retained via the first principal component (PC1) from principal component analysis (PCA). The resulting time series were filtered and transformed to power spectral density (PSD), allowing the calculation of average PSD amplitudes across the delta, theta, alpha, low-beta, high-beta, low-gamma, and high-gamma frequency bands. **c** Pain ratings were binarized to represent fluctuations on the basis of each annotation's median value.

Paired feature sets and pain ratings from each annotation formed the annotated pain series for model training.

---

**Wireless STN signals recording**

STN signal collection was obtained by activating BrainSense™ Survey via the Medtronic clinician programmer tablet [23] in the same visiting session as the VAS pain ratings and self-report collection.

BrainSense™ Survey records approximately 20-second time series signals with a 250 Hz sampling rate from the six possible electrode pairs (0-1, 1-2, 2-3, 0-2, 0-3, 1-3) on each side of the hemisphere (where '0' denotes the lowermost contact and '3' denotes the uppermost contact) with any stimulation is temporarily turned OFF. Patients were instructed to rest calmly with their eyes open in a seated, supine, or semisupine position, depending on their comfort. Patient signal data were downloaded anonymously in JavaScript Object Notation (JSON) format and transferred to a desktop PC for offline analysis. Rehabilitation therapy, medication, and stimulation were given on a patient basis for post-DBS dosage and stimulation adjustment and as part of hospital treatment.

**Self-reported symptoms**

To collect pain ratings and self-reported symptoms, participants used a touchscreen tablet (Apple iPad model A1980, Apple, Inc., Cupertino, California, USA) with customized software developed via the Swift programming language (v.4.1, Apple, Inc., Cupertino, California, USA) for the iOS system (v.11.4, Apple, Inc., Cupertino, California, USA). We consider sliding the visualized VAS scale on the touchscreen to be easier for patients than typing the values.

The self-reported symptoms selected for model development included mood and motor symptom assessments, which were based on the experience sampling method study recommendations.[50] The mood was assessed with four questions rated on a 1–7 VAS scale, addressing feelings of insecurity, relaxation, satisfaction, and feeling down. Five questions on Parkinson's disease (PD) motor symptoms were also included, each self-rated on a 1–7 VAS scale according to the suggested range. Fatigue was rated via a visual analogue scale (VAS) ranging from 0 to 100.[51] For each pain annotation, pain severity was assessed via a 0–100 VAS based on the patient's overall perception during visitation.[52,53] These questions and scales were determined through a focused group discussion among the authors (TY, NT, and AR). Supplementary Table S2 provides detailed descriptions of the self-reported items and ratings.

From the self-reported features, the motor symptoms model (MSY) included five features ("tremor," "rigidity," "walking difficulty," "imbalance," and "dyskinesia"). The mood-fatigue model (MOF) incorporated five features ("feeling down," "feeling insecure," "feeling relaxed," "feeling satisfied," and "fatigue") along with their respective ratings. Ratings were collected based on their overall perception during the visit.

**Software**

We utilized Python v.3.8.13,[54] and Python libraries to support data handling, analysis, model development, interpretation, and visualization. The libraries included Pandas v.1.4.2,[55] Numpy v.1.21.4,[56] Scipy  v.1.9.0,[57] Pingouin v.0.5.2,[58] Statsmodels v.0.13.2,[59] MLxtend v.0.19.0,[60] Scikit-learn v.1.0.2,[61] Shap v.0.39.0,[62] Matplotlib v.3.7.1,[63] and Seaborn v.0.11.2 and v0.13.2,[64] for data handling, analysis, model development, model interpretation, and visualization purposes, in this study.

**Preprocessing**

Python 3 (version 3.8.13) was employed primarily for computational analysis. Signals recorded via the BrainSense™ Survey from each patient's bilaterally implanted electrodes were prepared for model training. For each side, signals from three bipolar channel pairs (0-1, 1-2, and 2-3) with uniform 20-second signal lengths were selected for preprocessing.

To reduce dimensionality, principal component analysis (PCA) was applied to the three electrode pairs, and the first principal component (PC1) was retained as a single time series.[65] Subsequently, a 5th-order Butterworth bandpass filter with a high-pass frequency of 1 Hz and a low-pass frequency of 120 Hz was implemented via SciPy's signal module. The filter was applied forward and backward using "sosfiltfilt" as a neutral-phase filter.

For spectral analysis, the power spectral density (PSD) was calculated from filtered PC1 signals via Welch's method implemented in SciPy's signal module. This method employs a Hamming window of 250 points with 150 overlapping points, with the aforementioned sampling frequency and fast Fourier transform (FFT) for spectral estimation. Canonical band features were derived as the average of PSD amplitudes in the following frequency bands: delta (1–4 Hz), theta (5–8 Hz), alpha (9–12 Hz), low beta (13–20 Hz), high beta (21–30 Hz), low gamma (31–60 Hz), and high gamma (61–90 Hz). The STN-LFP model comprises fourteen non-normalized features based on the bilateral averaged canonical features described above.

The collected VAS ratings were converted into two binary class labels using the median value as a threshold, categorizing them as either "lower severity" or "higher severity." Observations at the median were assigned to lower or higher pain levels based on their proximity to minimum or maximum values, respectively. Annotated pain reports with distinct pain fluctuations represented by balanced binary distribution were used for model training. Specifically, imbalanced pain levels in which any single class label accounted for more than 60% of the observations[32] were excluded.

**Explainable machine learning**

The RF classifier is one of the most recommended classifier algorithms for various problems; it does not require input data normalization and can provide binary predictions and model explainability.[29,30] Nested cross-validation (CV) was employed with the RF classifier via the abovementioned features. The outer loop consisted of four repetitions of a 5-fold stratified CV to evaluate model performance on the basis of the best hyperparameters selected from the inner loop, ensuring that the training and test sets remain separated. The inner loop employed 5-fold stratified CV to select the best hyperparameters, which included [1, 2, 3, 4, 5, 6] as the minimum number of samples required at a leaf node, {"sqrt", None} as the number of features considered for optimal splitting, and "Gini" as the criterion for measuring split quality. The RF model and repeated-stratified 5-fold CV were controlled by setting the parameter *random-state* = 2022 to ensure reproducibility.

**Gini importance**

Model explainability to show how much a feature influences the predictions,[66] is achieved by calculating the mean decrease in impurity (MDI). This measure reflects the reduction in Gini impurity at each node split and is influenced by the features used to make that split.[29,66–69] In this study, Gini importance was computed via scikit-learn's RF implementation.[61] The feature importance scores were averaged across all repeated outer-loop cross-validations,[70] with higher Gini importance values indicating greater feature contributions to model performance.[68,71]

**SHAP values**

In addition to Gini importance, SHapley Additive exPlanations (SHAP) values[72] were used to provide post hoc interpretability for the RF model. SHAP offers directional values that enable both local and global interpretations, which can further complement the feature importance insights from Gini importance.

For this study, tree-based SHAP was applied to compute SHAP values from the RF model.[72,73] The model was retrained with the best hyperparameters identified during the nested cross-validation (CV) process[74] using the total observations.[75] In cases where multiple models achieved similar performance in the nested CV, the final set of hyperparameters was chosen following random selection among the best-performing models. Bar and violin plots were used to illustrate SHAP values, providing a detailed visualization of feature contributions.

**Bipolar electrode models**

Bipolar electrode models were developed by extracting the frequency-domain amplitude from each specific pair of electrodes, as provided by the BrainSense™ Survey,[23] via the Medtronic clinician programmer tablet. The power spectral density (PSD) amplitude was then averaged across previously defined canonical frequency bands (delta, theta, alpha, low-beta, high-beta, low-gamma, and high-gamma). These models were developed and evaluated via the same RF algorithm and nested CV scheme as those applied to the other models.

**Sample size**

The number of observations collected per patient varied according to hospitalization duration, patient availability, and patient's permission for data to be taken. Annotated pain reports with distinct pain fluctuations, represented by a balanced binary class,[32] and a minimum of ten observations per class were considered amenable for RF classifier training.[76] The total number of recruited patients was determined by the number of consenting patients who met the inclusion/exclusion criteria and the availability of hardware during the recruitment period.

**Statistics**

To identify whether pain ratings and time stamps when collecting data are associated, a scatter plot that visualizes the relationship between pain rating and the ANOVA procedure is used. The annotated pain series was divided into four groups (as the maximum number of observations a day), based on the 1st, 2nd, and 3rd quartile of the time stamps each day, to ensure similar distribution across the groups. One-way ANOVA was chosen to evaluate the difference across four groups of time. The significance level was set at $P < 0.05$. To compare the distribution of canonical bands' average PSD, a two-tailed Welch test is implemented with a significance level set at $P < 0.05$.

To evaluate the significance of model performance (balanced accuracy) for each annotated pain report, a permutation test was implemented by reshuffling the true labels 1000 times[77] while using the same model training approach described above. This process generated a distribution of 1000 balanced accuracy results expected by chance. Conservative estimates of $P$ values were calculated by dividing the occurrence of shuffled-label balanced accuracy values greater than or equal to the true-label balanced accuracy plus one by 1001.[77] The significance threshold for the permutation test was set at $P < 0.05$.

To evaluate group-level comparisons of true-label balanced accuracy against their corresponding chance-level performance, a one-sided Wilcoxon signed-rank test was applied by pairing each model's performance with its corresponding average of the 1000 distributions of shuffled-label balanced accuracy. The group-level alpha threshold was corrected to 0.0167 (0.05/3) to account for multiple comparisons of the STN-LFP, motor symptoms, mood, and fatigue models.

**Data availability**

The annotated pain reports for this manuscript are not publicly available because the data contain sensitive information that can be associated with specific persons. Requests to access the annotated pain reports and code utilized for the study should be directed to TY (tyanagisawa@nsurg.med.osaka-u.ac.jp) by the qualified researcher, who is subject to relevant ethical consideration.

**Code availability**

Requests to access the code utilized for the study should be directed to the corresponding author, TY (tyanagisawa@nsurg.med.osaka-u.ac.jp).

**Acknowledgements**

This research was funded by a grant from the Japan Agency for Medical Research and Development (AMED) (19dm0207070h0001). This work was also supported in part by the Japan Science and Technology Agency (JST), Core Research for Evolutional Science and Technology (JPMJCR18A5), the Exploratory Research for Advanced Technology (JPMJER1801), and the Moonshot R&D-MILLENNIA Program (JPMJMS2012), as well as Grants-in-Aid for Scientific Research from the Japan Society for the Promotion of Science (JSPS) (JP18H04085 and JP18H05522); AMED (JP19dm0307103 and JP19dm0307008)

We extend our heartfelt gratitude to all the patients, their families, and caregivers who supported this study. We would also like to thank He Xin (Ka Kaori), Keiko Inoshita, Ayaka Otsuka, and Yumiko Ogata for their exceptional administrative support, assistance with data collection, and effective communication with patients.

AR acknowledges the support of Grammarly, ChatGPT, and Perplexity.ai for improving readability and language fluency during the writing process. We acknowledge *www.fotor.com* for providing the AI-generated patient illustrations used in this manuscript.

**Author contributions**

Conceptualization: T.Y. and A.R.; methodology: A.R., N.T., and T.Y.; investigation: A.R., N.T., S.O., T.E., and T.Y.; data curation, formal analysis, visualizations: A.R.; data interpretation: A.R., and T.Y.; writing–original draft: A.R.; writing–review and editing: A.R., R.F., T.Y., B.S., N.T., S.O., T.E., and H.K.; supervision: T.Y., N.T., and H.K.; funding acquisition: T.Y., and H.K.; resource: H.K.

**Competing Interests**

The authors report no conflicts of interest.

**References**

1. Wasner, G. & Deuschl, G. Pains in Parkinson disease—many syndromes under one umbrella. *Nat Rev Neurol* **8**, 284–294 (2012).

2. Marques, A. & Brefel-Courbon, C. Chronic pain in Parkinson's disease: Clinical and pathophysiological aspects. *Revue Neurologique* **177**, 394–399 (2021).

3. Mostofi, A., Morgante, F., Edwards, M. J., Brown, P. & Pereira, E. A. C. Pain in Parkinson's disease and the role of the subthalamic nucleus. *Brain* **144**, 1342–1350 (2021).

4. Mylius, V. *et al.* Pain in Parkinson's Disease: Current Concepts and a New Diagnostic Algorithm. *Mov Disord Clin Pract* **2**, 357–364 (2015).

5. Lee, M. A., Walker, R. W., Hildreth, T. J. & Prentice, W. M. A Survey of Pain in Idiopathic Parkinson's Disease. *Journal of Pain and Symptom Management* **32**, 462–469 (2006).

6. Broen, M. P. G., Braaksma, M. M., Patijn, J. & Weber, W. E. J. Prevalence of pain in Parkinson's disease: A systematic review using the modified QUADAS tool. *Movement Disorders* **27**, 480–484 (2012).

7. Kurihara, K. *et al.* Fluctuating pain in Parkinson's disease: Its prevalence and impact on quality of life. *eNeurologicalSci* **25**, 100371 (2021).

8. Quinn, N. P. Classification of fluctuations in patients with Parkinson's disease. *Neurology* **51**, (1998).

9. Samura, K. *et al.* Intractable facial pain in advanced Parkinson's disease alleviated by subthalamic nucleus stimulation. *Journal of Neurology, Neurosurgery & Psychiatry* **79**, 1410–1411 (2008).

10. Storch, A. *et al.* Pain Fluctuations in Parkinson's Disease and Their Association with Motor and Non-Motor Fluctuations. *JPD* 1–18 (2024) doi:10.3233/JPD-240026.

11. Storch, A. *et al.* Nonmotor fluctuations in Parkinson disease: Severity and correlation with motor complications. *Neurology* **80**, 800–809 (2013).

12. Desch, S., Schweinhardt, P., Seymour, B., Flor, H. & Becker, S. Evidence for dopaminergic involvement in endogenous modulation of pain relief. *eLife* **12**, e81436 (2023).

13. Bannister, K., Bee, L. A. & Dickenson, A. H. Preclinical and Early Clinical Investigations Related to Monoaminergic Pain Modulation. *Neurotherapeutics* **6**, 703–712 (2009).

14. Giannicola, G. *et al.* The effects of levodopa and ongoing deep brain stimulation on subthalamic beta oscillations in Parkinson's disease. *Experimental Neurology* **226**, 120–127 (2010).

15. Mathiopoulou, V. *et al.* Modulation of subthalamic beta oscillations by movement, dopamine, and deep brain stimulation in Parkinson's disease. *npj Parkinsons Dis.* **10**, 1–7 (2024).

16. Lofredi, R. *et al.* Dopamine-dependent scaling of subthalamic gamma bursts with movement velocity in patients with Parkinson's disease. *eLife* **7**, e31895 (2018).

17. Sand, D. *et al.* Machine learning-based personalized subthalamic biomarkers predict ON-OFF levodopa states in Parkinson patients. *J. Neural Eng.* **18**, 046058 (2021).

18. Morelli, N. & Summers, R. L. S. Association of subthalamic beta frequency sub-bands to symptom severity in patients with Parkinson's disease: A systematic review. *Parkinsonism & Related Disorders* **110**, 105364 (2023).

19. Belasen, A. *et al.* The Effects of Mechanical and Thermal Stimuli on Local Field Potentials and Single Unit Activity in Parkinson's Disease Patients. *Neuromodulation: Technology at the Neural Interface* **19**, 698–707 (2016).

20. Parker, T. *et al.* Pain-Induced Beta Activity in the Subthalamic Nucleus of Parkinson's Disease. *Stereotact Funct Neurosurg* **98**, 193–199 (2020).

21. Diao, Y. *et al.* A Meta-Analysis of the Effect of Subthalamic Nucleus-Deep Brain Stimulation in Parkinson's Disease-Related Pain. *Frontiers in Human Neuroscience* **15**, (2021).

22. Little, S. & Brown, P. Debugging Adaptive Deep Brain Stimulation for Parkinson's Disease. *Movement Disorders* **35**, 555–561 (2020).

23. Jimenez-Shahed, J. Device profile of the percept PC deep brain stimulation system for the treatment of Parkinson's disease and related disorders. *Expert Review of Medical Devices* **18**, 319–332 (2021).

24. Constantoyannis, C., Berk, C., Honey, C. R., Mendez, I. & Brownstone, R. M. Reducing Hardware-Related Complications of Deep Brain Stimulation. *Canadian Journal of Neurological Sciences* **32**, 194–200 (2005).

25. Feldmann, L. K. *et al.* Toward therapeutic electrophysiology: beta-band suppression as a biomarker in chronic local field potential recordings. *npj Parkinsons Dis.* **8**, 44 (2022).

26. Shirvalkar, P. *et al.* First-in-human prediction of chronic pain state using intracranial neural biomarkers. *Nat Neurosci* **26**, 1090–1099 (2023).

27. Naeini, E. K. *et al.* Pain Recognition With Electrocardiographic Features in Postoperative Patients: Method Validation Study. *Journal of Medical Internet Research* **23**, e25079 (2021).

28. Thompson, J. A., Lanctin, D., Ince, N. F. & Abosch, A. Clinical Implications of Local Field Potentials for Understanding and Treating Movement Disorders. *Stereotact Funct Neurosurg* **92**, 251–263 (2014).

29. Archer, K. J. & Kimes, R. V. Empirical characterization of random forest variable importance measures. *Computational Statistics & Data Analysis* **52**, 2249–2260 (2008).

30. Chicco, D. & Oneto, L. Computational intelligence identifies alkaline phosphatase (ALP), alpha-fetoprotein (AFP), and hemoglobin levels as most predictive survival factors for hepatocellular carcinoma. *Health Informatics J* **27**, 1460458220984205 (2021).

31. Mayr, A. *et al.* Patients with chronic pain exhibit individually unique cortical signatures of pain encoding. *Human Brain Mapping* **43**, 1676–1693 (2022).

32. Wei, Q. & Dunbrack Jr, R. L. The role of balanced training and testing data sets for binary classifiers in bioinformatics. *PloS one* **8**, e67863 (2013).

33. Margolis, R. B., Tait, R. C. & Krause, S. J. A rating system for use with patient pain drawings: *Pain* **24**, 57–65 (1986).

34. Benerradi, J., Clos, J., Landowska, A., Valstar, M. F. & Wilson, M. L. Benchmarking framework for machine learning classification from fNIRS data. *Frontiers in Neuroergonomics* **4**, (2023).

35. Sürücü, O., Baumann-Vogel, H., Uhl, M., Imbach, L. L. & Baumann, C. R. Subthalamic deep brain stimulation versus best medical therapy for l-dopa responsive pain in Parkinson's disease. *Pain* **154**, 1477–1479 (2013).

36. Cury, R. G. *et al.* Subthalamic deep brain stimulation modulates conscious perception of sensory function in Parkinson's disease. *Pain* **157**, 2758–2765 (2016).

37. Marques, A. *et al.* Central pain modulation after subthalamic nucleus stimulation: A crossover randomized trial. *Neurology* **81**, 633–640 (2013).

38. Dogru Huzmeli, E., Yilmaz, A. & Okuyucu, E. Analysis of the effects of subthalamic nucleus deep brain stimulation on somatosensation in Parkinson's disease patients. *Neurol Sci* **41**, 925–931 (2020).

39. Redding, A. *et al.* Medically unexplained pain and suicidal ideation among US adults. *Journal of Affective Disorders* **351**, 425–429 (2024).

40. Ruben, J. Somatotopic Organization of Human Secondary Somatosensory Cortex. *Cerebral Cortex* **11**, 463–473 (2001).

41. Shreve, L. A. *et al.* Subthalamic oscillations and phase amplitude coupling are greater in the more affected hemisphere in Parkinson's disease. *Clinical Neurophysiology* **128**, 128–137 (2017).

42. Tajerian, M., Amrami, M. & Betancourt, J. M. Is there hemispheric specialization in the chronic pain brain? *Experimental Neurology* **355**, 114137 (2022).

43. Darvas, F. & Hebb, A. O. Task specific inter-hemispheric coupling in human subthalamic nuclei. *Front. Hum. Neurosci.* **8**, (2014).

44. Arnold Anteraper, S. *et al.* Resting-State Functional Connectivity of the Subthalamic Nucleus to Limbic, Associative, and Motor Networks. *Brain Connectivity* **8**, 22–32 (2018).

45. Prenassi, M. *et al.* Peri-lead edema and local field potential correlation in post-surgery subthalamic nucleus deep brain stimulation patients. *Front. Hum. Neurosci.* **16**, (2022).

46. von Elm, E. *et al.* The Strengthening the Reporting of Observational Studies in Epidemiology (STROBE) statement: guidelines for reporting observational studies. *The Lancet* **370**, 1453–1457 (2007).

47. Maruo T. *et al.* Translation and reliability and validity of a Japanese version of the revised Short-Form McGill Pain Questionnaire (SF-MPQ-2). *PAIN RESEARCH* **28**, 43–53 (2013).

48. Okuma, Y. *et al.* Fatigue in Japanese patients with Parkinson's disease: A study using Parkinson fatigue scale. *Movement Disorders* **24**, 1977–1983 (2009).

49. Tani, N., Yaegaki, T. & Kishima, H. A Case Report: Hemorrhagic Venous Infarction after Deep Brain Stimulation Surgery Probably Due to Coagulation of Intradural Veins. *NMC Case Report Journal* **8**, 315–318 (2021).

50. Mulders, A. E. P. *et al.* Usability of the Experience Sampling Method in Parkinson's Disease on a Group and Individual Level. *Movement Disorders* **35**, 1145–1152 (2020).

51. Jason, L. A. *et al.* Contrasting Case Definitions for Chronic Fatigue Syndrome, Myalgic Encephalomyelitis/Chronic Fatigue Syndrome and Myalgic Encephalomyelitis. *Eval Health Prof* **35**, 280–304 (2012).

52. Jensen, M. Interpretation of visual analog scale ratings and change scores: a reanalysis of two clinical trials of postoperative pain. *The Journal of Pain* **4**, 407–414 (2003).

53. Kelly, A. Does the Clinically Significant Difference in Visual Analog Scale Pain Scores Vary with Gender, Age, or Cause of Pain? *Academic Emergency Medicine* **5**, 1086–1090 (1998).

54. Van Rossum, G. & Drake, F. Python 3 Reference Manual. Scotts Valley, CA: CreateSpace; 2009. (2020).

55. McKinney, W. Data Structures for Statistical Computing in Python. in 56–61 (Austin, Texas, 2010). doi:10.25080/Majora-92bf1922-00a.

56. Harris, C. R. *et al.* Array programming with NumPy. *Nature* **585**, 357–362 (2020).

57. Virtanen, P. *et al.* SciPy 1.0: fundamental algorithms for scientific computing in Python. *Nat Methods* **17**, 261–272 (2020).

58. Vallat, R. Pingouin: statistics in Python. *Journal of Open Source Software* **3**, 1026 (2018).

59. Seabold, S. & Perktold, J. Statsmodels: Econometric and Statistical Modeling with Python. in 92–96 (Austin, Texas, 2010). doi:10.25080/Majora-92bf1922-011.

60. Raschka, S. MLxtend: Providing machine learning and data science utilities and extensions to Python's scientific computing stack. *Journal of Open Source Software* **3**, 638 (2018).

61. Pedregosa, F. *et al.* Scikit-learn: Machine learning in Python. *the Journal of machine Learning research* **12**, 2825–2830 (2011).

62. Lundberg, S. M. & Lee, S.-I. A Unified Approach to Interpreting Model Predictions. in *Advances in Neural Information Processing Systems* vol. 30 (Curran Associates, Inc., 2017).

63. Hunter, J. D. Matplotlib: A 2D Graphics Environment. *Computing in Science & Engineering* **9**, 90–95 (2007).

64. Waskom, M. L. seaborn: statistical data visualization. *Journal of Open Source Software* **6**, 3021 (2021).

65. Asadur Rahman, Md., Foisal Hossain, Md., Hossain, M. & Ahmmed, R. Employing PCA and t-statistical approach for feature extraction and classification of emotion from multichannel EEG signal. *Egyptian Informatics Journal* **21**, 23–35 (2020).

66. Mecikalski, J. R. *et al.* A Random-Forest Model to Assess Predictor Importance and Nowcast Severe Storms Using High-Resolution Radar–GOES Satellite–Lightning Observations. *Monthly Weather Review* **149**, 1725–1746 (2021).

67. Loecher, M. Unbiased variable importance for random forests. *Communications in Statistics - Theory and Methods* **51**, 1413–1425 (2022).

68. Saarela, M. & Jauhiainen, S. Comparison of feature importance measures as explanations for classification models. *SN Appl. Sci.* **3**, 272 (2021).

69. Ward, I. R. *et al.* Explainable artificial intelligence for pharmacovigilance: What features are important when predicting adverse outcomes? *Computer Methods and Programs in Biomedicine* **212**, 106415 (2021).

70. Mitra, J. *et al.* Statistical machine learning to identify traumatic brain injury (TBI) from structural disconnections of white matter networks. *NeuroImage* **129**, 247–259 (2016).

71. Hanuschkin, A. *et al.* Investigation of cycle-to-cycle variations in a spark-ignition engine based on a machine learning analysis of the early flame kernel. *Proceedings of the Combustion Institute* **38**, 5751–5759 (2021).

72. Lundberg, S. M. *et al.* From local explanations to global understanding with explainable AI for trees. *Nat Mach Intell* **2**, 56–67 (2020).

73. Obaido, G. *et al.* An Interpretable Machine Learning Approach for Hepatitis B Diagnosis. *Applied Sciences* **12**, 11127 (2022).

74. Raschka, S. Model Evaluation, Model Selection, and Algorithm Selection in Machine Learning. Preprint at http://arxiv.org/abs/1811.12808 (2020).

75. Schletz, D., Breidung, M. & Fery, A. Validating and Utilizing Machine Learning Methods to Investigate the Impacts of Synthesis Parameters in Gold Nanoparticle Synthesis. *J. Phys. Chem. C* **127**, 1117–1125 (2023).

76. Martha, S. R., Chen, K.-F., Lin, Y. & Thompson, H. J. Plasma Phospholipid Metabolites Associate With Functional Outcomes Following Mild Traumatic Brain Injury in Older Adults. *Biological Research For Nursing* **23**, 127–135 (2021).

77. Vieira, S., Garcia-Dias, R. & Lopez Pinaya, W. H. Chapter 19 - A step-by-step tutorial on how to build a machine learning model. in *Machine Learning* (eds. Mechelli, A. & Vieira, S.) 343–370 (Academic Press, Cambridge, United Kingdom, 2020). doi:10.1016/B978-0-12-815739-8.00019-5.

**Supplementary information for:**

**Wirelessly transmitted subthalamic nucleus signals could predict endogenous pain fluctuation levels in Parkinson's disease patients**

Abdi Reza[1,2], Takufumi Yanagisawa[1,3] *, Naoki Tani[1] **, Ryohei Fukuma[1,3], Takuto Emura[1], Satoru Oshino[1], Ben Seymour[4,5], Haruhiko Kishima[1]

[1] Department of Neurosurgery, Graduate School of Medicine, Osaka University, Osaka 565-0871, Japan.

[2] Department of Neurosurgery, Faculty of Medicine, Universitas Indonesia, Jakarta 10430, Jakarta, Indonesia.

[3] Institute for Advanced Co-Creation Studies, Osaka University, Suita, Osaka 565-0871, Japan.

[4] Institute of Biomedical Engineering, University of Oxford, Oxford OX37DQ, England, UK.

[5] Wellcome Centre for Integrative Neuroimaging, University of Oxford, Oxford OX39DU, England, UK.

**SUPPLEMENTARY RESULTS**

**Report-specific patterns in binary classes of pain experience**

The STN-LFP, mood, and fatigue features exhibited variability across annotated pain reports regarding the features associated with higher and lower pain experiences (Fig. 4). However, the motor symptoms feature showed greater pain levels corresponding to a higher average of rigidity and imbalance across annotated pain reports (Supplementary Fig. S9).

**Mood and fatigue feature prediction for binary-class pain**

The mood and fatigue features yielded significant decoding performance in three of the six reports with PDRP characteristics consisting of (average of balanced accuracy ($P$value, permutation test)): 68.79 (0.007), 67.38(0.127), 86.46(0.001), for reports #A, #B, and #C, respectively as indicated by significant intraindividual models ($P < 0.05$, permutation test; Supplementary Fig. S10, S11; Table S4). Group-level analysis of the six annotated pain with PDRP characteristics revealed a mean balanced accuracy of 69.71% $\pm$ 8.40 (mean $\pm$ SD), which was significantly greater than the corresponding average of shuffled data ($P = 0.0156$, n = 6, one-sided Wilcoxon signed-rank test, with a Bonferroni-adjusted α-level of 0.0167 [0.05/3]; Supplementary Fig. S10, S11; Table S4). The mood and fatigue features yielded significant decoding performance in one of the two reports with non-PDRP characteristics consisting of (average of balanced accuracy ($P$value, permutation test)): 67.80% (0.004), for report #H (Supplementary Fig. S10, S11; Table S4).

**Motor symptom feature prediction for binary-class pain**

The motor symptoms features yielded significant decoding performance in four of the six reports with PDRP characteristics consisting of (average of balanced accuracy ($P$value, permutation test)): 79.25(0.002), 76.88(0.002), 74.17(0.015), 74.38(0.010) for reports #B, #C, #E, and #F respectively as indicated by significant intraindividual models ($P < 0.05$, permutation test; Supplementary Fig. S10, S11; Table S4). Group-level analysis of the six annotated pain with PDRP characteristics revealed a mean balanced accuracy of 67.66% $\pm$ 14.92 (mean $\pm$ SD), which was not significantly higher than the corresponding chance level after Bonferroni correction but was still lower than 0.05 ($P = 0.031$, n = 6, one-sided Wilcoxon signed-rank test, with a Bonferroni-adjusted α-level of 0.0167 [0.05/3]; Supplementary Fig. S10, S11; Table S4). The motor symptoms features yielded significant decoding performance in one of the two reports with non-PDRP characteristics consisting of (average of balanced accuracy ($P$value, permutation test)): 62.26 (0.043) for report #H (Supplementary Fig. S10, S11; Table S4).

**Model explainability**

Our analysis of bipolar electrode models revealed various performances on the basis of the side of the subthalamic nucleus (STN) and specific bipolar electrode configurations (Supplementary Fig. S12). The electrode-specific model of reports #B, #C, #E, and #F shows average performance above 50% chance level with better trend in contralateral electrodes for #B, #E, and right electrodes for #C, #F in accordance with STN-LFP model and model explainability about STN with dominant influence. Similarly, reports #A, #D, #G, and #H that have nonsignificant decoding performance show electrode-specific performance distributed around 50% chance levels.

The descriptions of the Gini importance and SHAP values are reported in Fig. S14-21. The Gini importance of "fatigue" and the "feeling down" feature presented the highest median values across annotated pain reports #A, #B, #C, and #H. Similarly, "imbalance" and "rigidity" presented the highest median values across annotated pain reports #B, #C, #E, and #H (Fig. S22).

**SUPPLEMENTARY DISCUSSION**

**Different bipolar pairs lead to varying performance**

Our study revealed variability in model performance with different bipolar electrode pairs for predicting pain in PD patients. Electrode selection for STN analysis is often based on the beta band,[1,2] therapeutic stimulation targets[3,4] or anatomical location of the electrode.[5] However, these approaches typically focus on the influence of the beta band on motor symptoms. In contrast, we implemented PCA for dimensionality reduction across three pairs of electrodes for the main LFP model without presuming which frequency band is linked explicitly to pain. Our findings suggest that this approach can effectively decode binary pain experiences (Fig. 5, Supplementary Fig. S10, and Table S4). Nevertheless, determining the optimal electrode pair for pain decoding remains essential for practical implementation.

Bipolar electrode models help identify which electrode pairs provide optimal decoding performance for practical applications. The bipolar electrode model demonstrated varying performance across STN sides and specific electrode pairs. This variability in performance across different STN regions and electrode pairs suggests that the relationship between LFP and pain experiences in PD patients, other than the canonical band's influence, may also be spatially influenced inside the STN.

**Mood and fatigue features can complement the STN-LFP model**

Mood, fatigue, and motor symptom features were used to train alternative models to STN-LFP. Both mood[6] and fatigue[7,8] are closely associated with pain experiences. Shared mechanisms between mood, fatigue, and pain have been recognized, particularly through glutamatergic overactivity.[9,10] Fatigue may exacerbate pain, whereas pain disrupts sleep,[11] thus creating a vicious cycle.[12] Moreover, mood plays a significant role in pain modulation via the lateral inferior frontal cortex[13] and mood disorders have been associated with pain experience.[14]

Our findings show that the mood and fatigue model can complement the STN-LFP model. Specifically, the mood and fatigue model provided significant performance in two annotated pain reports where LFP features also led to significant results (#B and #C), in one annotated pain report where only mood and fatigue models provided significant results (#A), and for a non-PDRP annotated pain report (#H). This finding suggests that the mood and fatigue model offers additional benefits under certain conditions and could serve as a valuable complement to the LFP model.

**The STN-LFP model is comparable to the motor symptoms model**

Our study revealed that the STN-LFP model performs comparably to the motor symptoms model in predicting pain in PD patients. This finding aligns with established associations between pain and motor symptoms in PD, including dyskinesia,[15] rigidity,[16] PD motor fluctuations,[17] which can affect pain. Unsurprisingly, STN-LFP shows the association between specific frequency bands and motor symptom types, including the low-beta band, which is associated with rigidity, bradykinesia, and freezing of gait, and the gamma band, which is associated with dyskinesia.[18]  In the long term, pathologically increased muscle tone could increase the frequency of pain,[19] and initiate other musculoskeletal disorders, thus increasing the degree of pain.[19–21]

In our analysis of six PDRP annotated pain reports, motor symptoms provided above-chance performance in four annotated pain reports (#B, #C, #E, and #F), overlapping with annotated pain reports where the STN-LFP model showed similar efficacy. Notably, the STN-LFP model offers an advantage because it relies on objective, electrophysiological measurements, potentially providing more reliable and precise indicators of pain experience than subjective motor symptom assessments alone.

PD symptoms can influence pain, even when its origin is not directly related to the disease, and this pain may be reflected in self-reports rather than in STN signals. This could indicate that in addition to STN recordings,[22] the motor activity representations may require other electrophysiological recordings, such as thalamic signals.[23] In one such annotated pain report (#H), motor symptom features demonstrated significant predictive performance for pain experiences.

The ability of motor symptom ratings to identify pain in this context suggests that PD's influence on pain extends beyond direct causation, suggesting that motor symptoms can play a role in identifying pain even when its origin is not directly related to PD pathology. This observation aligns with the concept of "PD indirectly aggravated pain" or "pain indirectly related to idiopathic PD,"[24–26] where PD exacerbates the pain of non-PD origin; such indirectly related pain can be more challenging to manage than PD-related pain.[25]

Two annotated pain reports (D and G) did not lead to significant model performance with any features. This observation suggests that pain occurrence cannot be accurately predicted in some cases via objective measurements (such as STN-LFP) or subjective measures (such as self-reported data). Further investigations into the origins, causes, and classifications of pain will be crucial to guide more effective treatment plans.

**The benefit of the self-report model**

The Gini importance and SHAP values can help sort and quantify the influence of self-reported features, providing valuable insights into how specific symptoms, such as fatigue or mood, contribute to pain experiences. Identifying key motor symptoms and other self-reported factors allows physicians to prioritize interventions that target these symptoms to alleviate pain.[27] Interventions targeting fatigue, such as tailored exercise programs, sleep management strategies, and appropriate medication adjustments, may indirectly benefit pain management.[28,29] Moreover, the self-reported model's insights into mood and its relationship with pain can guide personalized treatment plans. For selected patients, this may involve pharmacological interventions targeting mood disorders [9] or referrals for psychological counseling and cognitive behavioral approaches.[30]

## Supplementary Figure S1. Excluded annotated pain reports

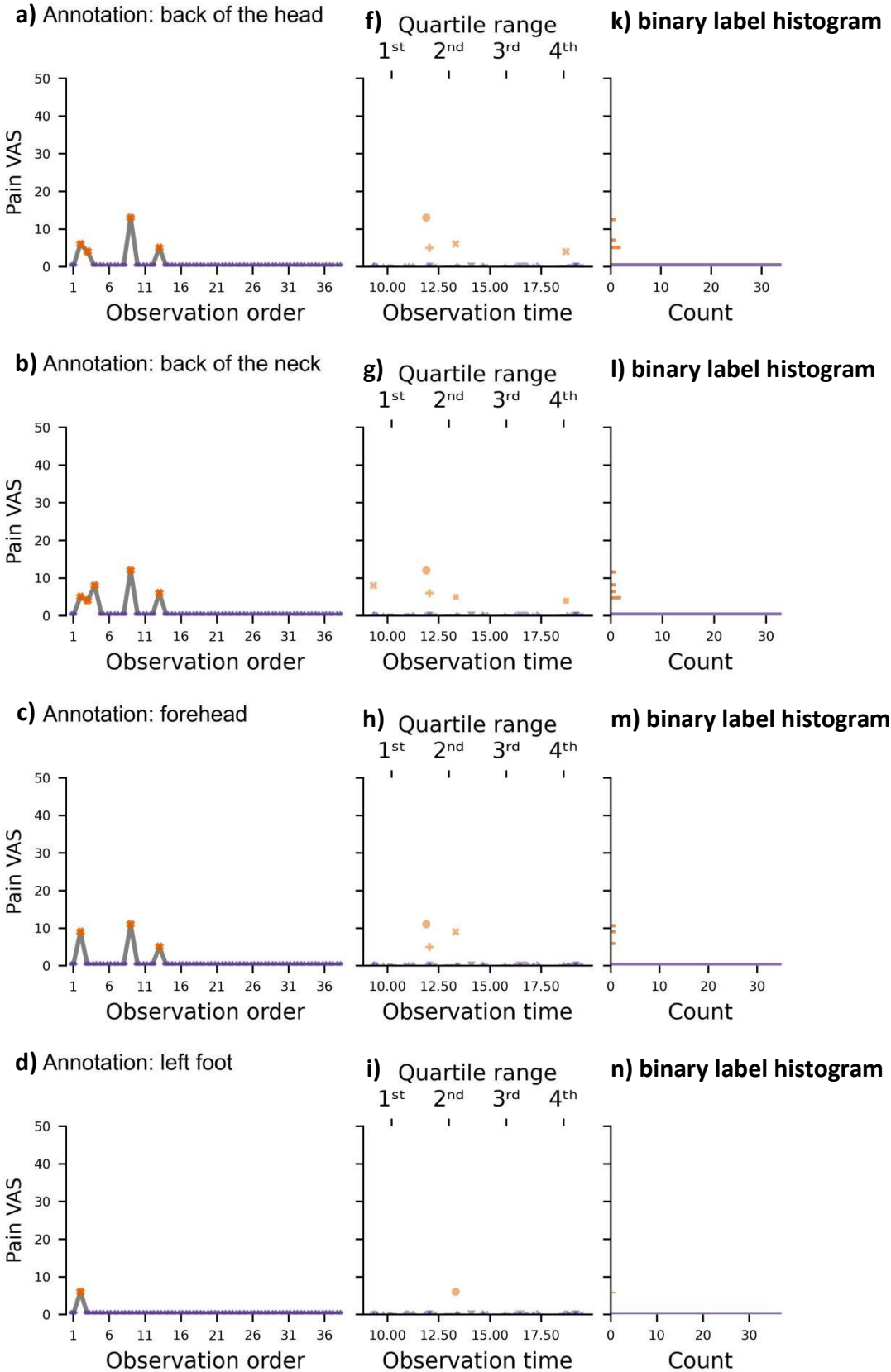

**Supplementary Figure S1.** Schematic representation of pain series from one female patient, 55 years old, who reported five pain annotations before surgery (a-e). Post-DBS system implantation, annotated pain reports were followed with VAS ratings (a-e: f-j), wireless STN signals transmission, and other self-reports. Warm colors represent higher pain levels; cold colors represent lower pain levels.

Pain transformation showed imbalanced binary label distribution after median splitting: **k** pain annotation "back of the head" has (amount of observations (min-max)) 34(0-0) for lower and 4(4-13) for higher pain labels, **l** pain annotation "back of the neck" has 33(0-0) for lower and 5(4-12) for higher pain labels, **m** pain annotation "forehead" has 35(0-0) for lower and 3(5-11) for higher pain labels,  **n** pain annotation "left foot" has 37(0-0) for lower and 1(6-6) for higher pain labels. These are representatives of annotated pain series identified as 'lack of fluctuations' or 'non-fluctuations' reports. Patient and pain reports were therefore excluded from further model training.

## Supplementary Figure S2. Report A

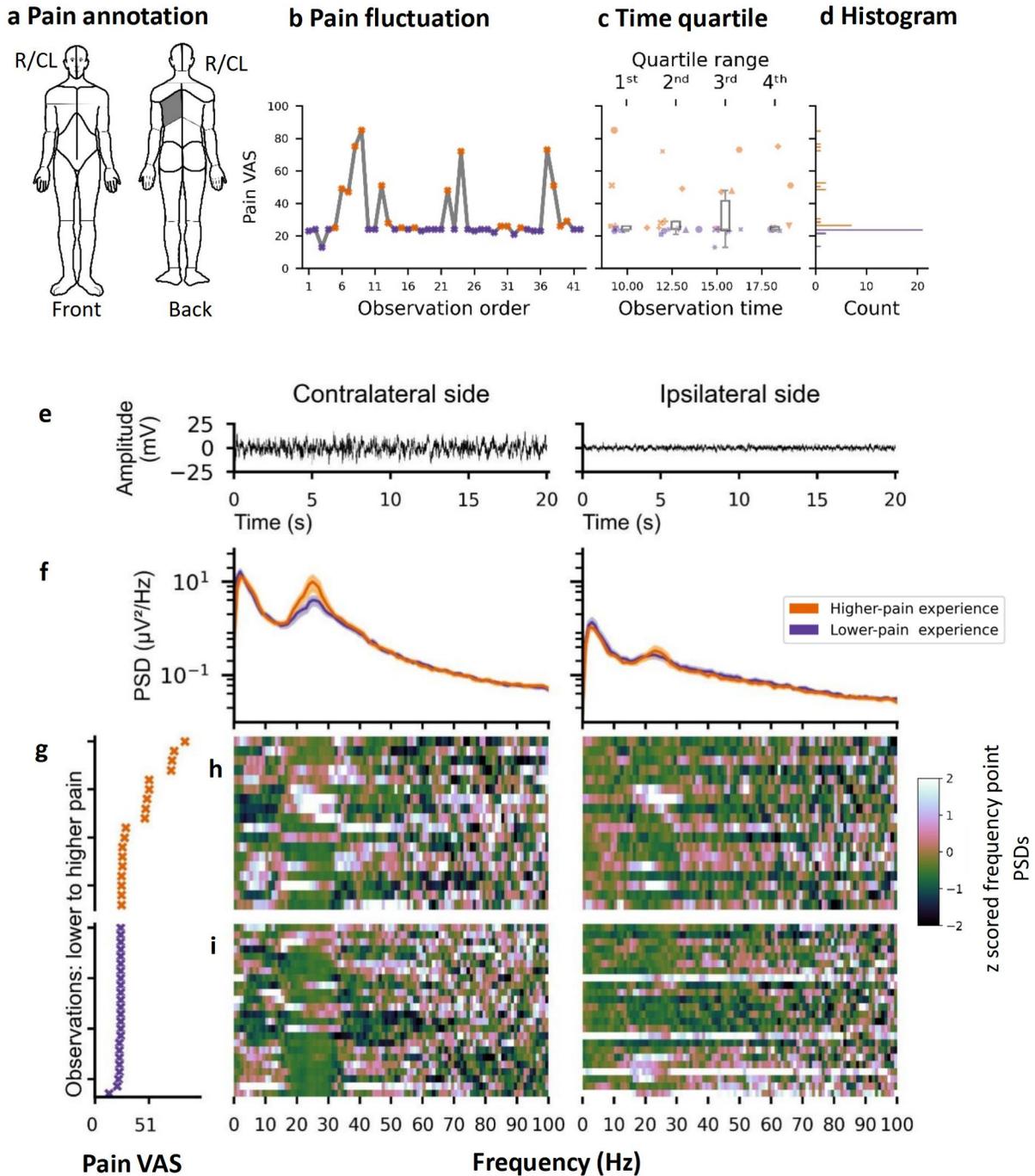

**Figure S2.** Individual pain fluctuations in annotated pain report #A. **a** Pain map based on patient-reported pain locations, with an annotation focused on the left back area. The schematic body pain area representation is adapted from Margolis et al.[31] The grey shading indicates areas of reported pain (R: right; CL: contralateral to the pain location). **b** Pain fluctuations over the observation period were plotted sequentially. **c** Scatterplot showing individual pain ratings by daily timestamp. Data points recorded on the same day are represented by matching symbols. Boxplots overlay the scatterplot, illustrating the central tendency and distribution within four quartile divisions of the annotated pain report timestamps: median (central line), first and third quartiles (box borders), and whiskers extending to 1.5 times the interquartile range. No significant differences were observed in pain distribution across quartiles (Annotated pain report A; $F(41) = 0.08$, $P = 0.999$; one-way ANOVA). **d** Histogram displaying the distribution of transformed pain ratings, divided into higher-pain and lower-pain classes. **e** STN time series signals correspond to the recording site. **f** Average power spectral density (PSD) amplitudes for higher- and lower-pain experiences are shown on a log scale for the y-axis, with shaded areas indicating the standard error of the mean (SEM). **g** Observations reordered from lower to higher pain ratings. (h, i) Heatmaps of reordered signal recordings, with z-scored PSD values for higher-pain **h** and lower-pain **i** levels within the same frequency range.

## Supplementary Figure S3. Report B

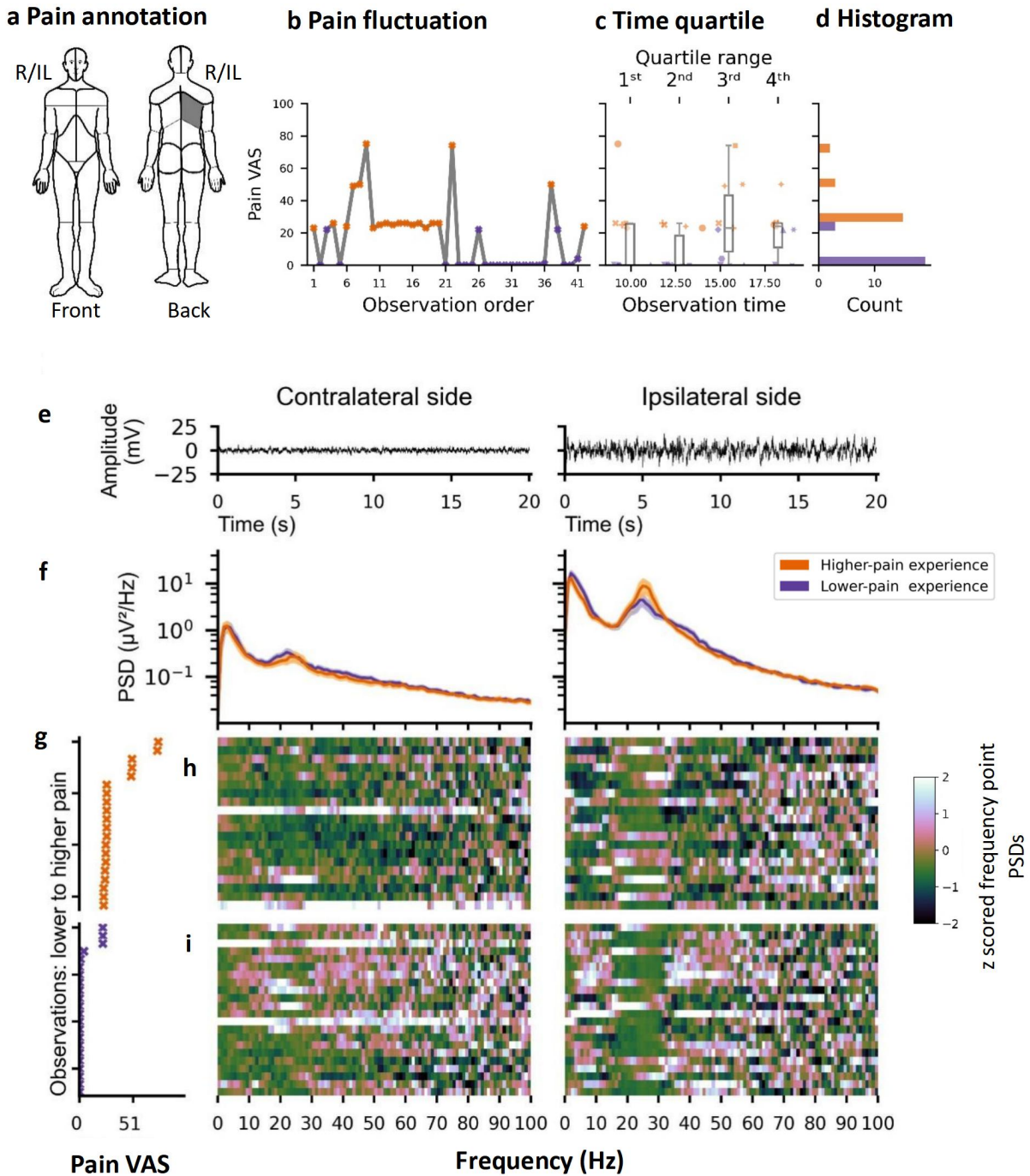

**Figure S3.** Individual pain fluctuations in annotated pain report #B. **a** Pain map based on patient-reported pain locations, with an annotation in the right back area. The schematic body pain area representation is adapted from Margolis et al.[31] The grey shading indicates reported pain locations (R: right; IL: ipsilateral to pain location). **b** Pain fluctuation over the observation sequence. **c** Scatterplot of individual pain ratings by daily timestamp; data points recorded on the same day are represented with matching symbols. Boxplots overlay the scatterplot, showing the central tendency and distribution across four quartile divisions of annotated pain report timestamps: median (central line), first and third quartiles (box borders), and whiskers extending to 1.5 times the interquartile range. No significant differences in pain distribution were observed across quartiles (Annotated pain report B; $F(41) = 1.79$, $P = 0.17$; one-way ANOVA). **d** Histogram displaying transformed pain ratings, divided into higher-pain and lower-pain classes. **e** STN time series signals correspond to recording site. **f** Average power spectral density (PSD) amplitudes for higher- and lower-pain experiences are shown on a log scale for the y-axis, with shaded areas indicating the standard error of the mean (SEM). **g** Observations reordered from lower to higher pain ratings. **h,i** Heatmaps of reordered signal recordings, with z scored PSD values for higher-pain **h** and lower-pain **i** levels within the same frequency range

## Supplementary Figure S4. Report C

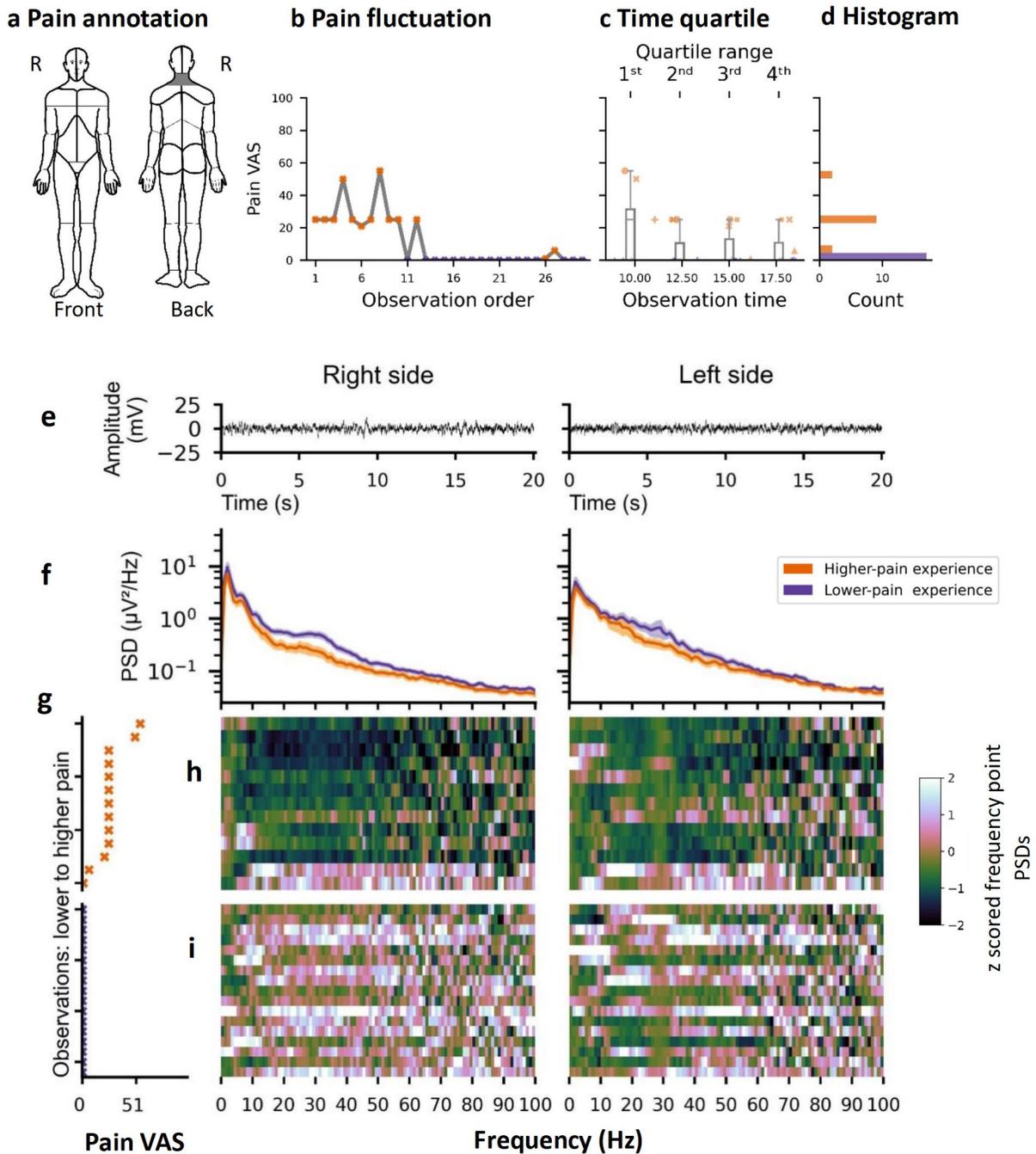

**Figure S4.** Individual pain fluctuations in annotated pain report #C. **a** Pain map illustrating the pain location reported by the patient, showing annotations in the back of the neck area. The schematic body pain area representation is adapted from Margolis et al.[31] The grey shading indicates the reported pain location (R: right). **b** Pain fluctuations across the observation sequence. **c** Scatterplot of individual pain ratings, with the daily timestamp of recordings; symbols represent data recorded on the same day. The overlayed boxplots depict central tendencies and data distributions across four quartile ranges of the annotated pain report timestamp: median (central line), first and third quartiles (box borders), and whiskers extending to 1.5 times the interquartile range. No significant difference was found across quartiles for pain reports (Annotated pain report C; $F(29) = 2.09$, $P = 0.13$; one-way ANOVA). **d** Histogram of transformed pain ratings divided into higher-pain and lower-pain classes. **e** STN time series signals correspond to the recording site. **f** Average power spectral density (PSD) amplitudes for higher- and lower-pain experiences are shown on a log scale for the y-axis, with shaded areas indicating the standard error of the mean (SEM). **g** Observations reordered from lower to higher pain ratings. **h,i** Heatmaps of reordered signal recordings, with z scored PSD values for higher-pain **h** and lower-pain **i** levels within the same frequency range.

**Supplementary Figure S5. Report D**

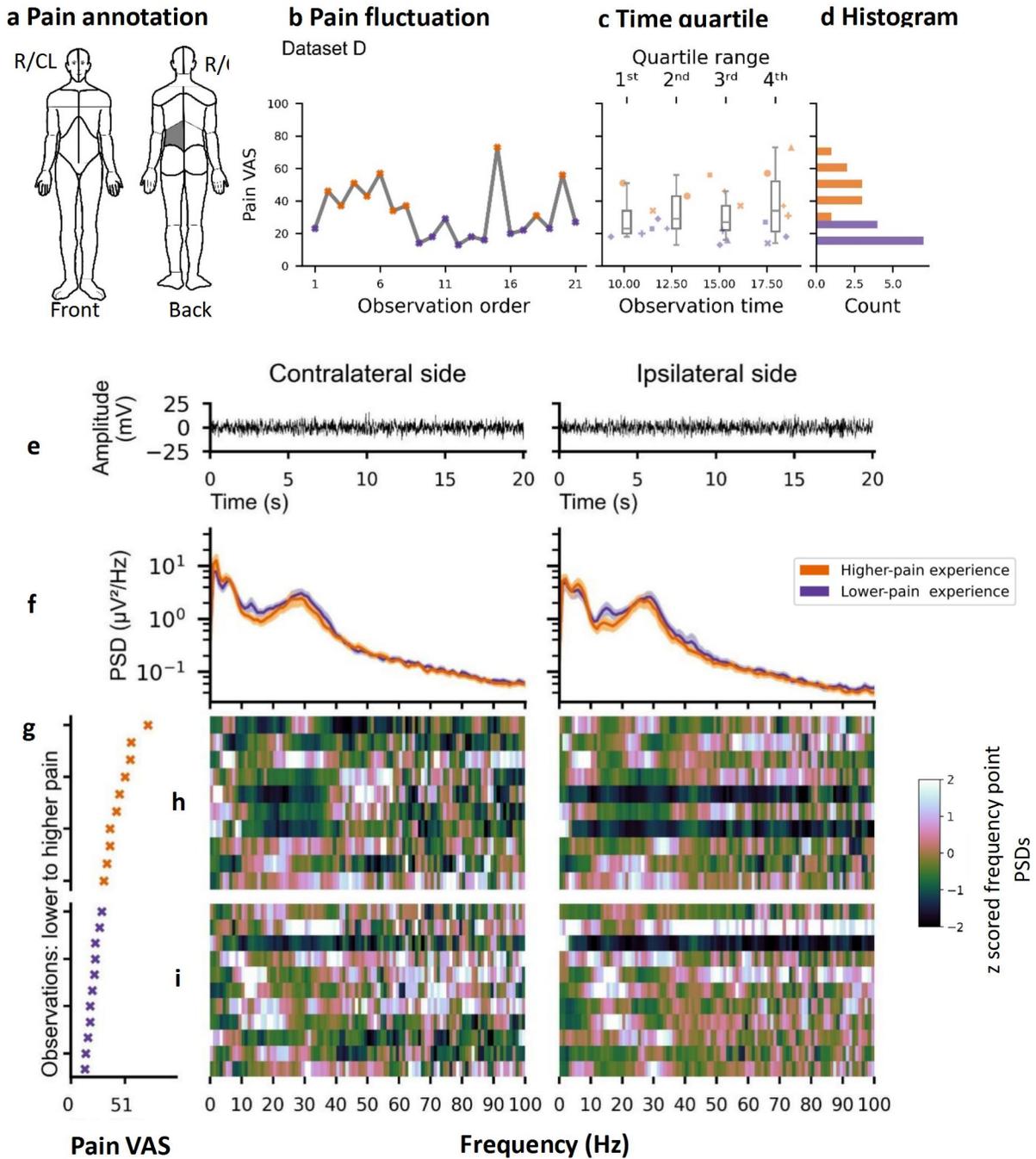

**Figure S5.** Individual pain fluctuations in annotated pain report #D. **a** Pain map based on the patient's report, showing pain annotations in the left lower back area. The schematic body pain representation is adapted from Margolis et al.[31] The grey shading represents pain annotation (R: right; CL: contralateral side of the pain annotation). **b** Pain fluctuations across the observation sequence. **c** Scatterplot displaying individual pain ratings alongside the daily timestamp of data recordings, with same-day recordings marked by identical symbols. Boxplots overlay the scatterplot to show central

tendencies and distributions across four timestamp-based quartiles: medians (centrelines), first and third quartiles (box borders), and whiskers extending to 1.5 times the interquartile range. No significant difference was observed in pain distribution across the quartiles (Annotated pain report D; $F(20) = 0.33$, $P = 0.80$; one-way ANOVA). **d** Histogram showing the distribution of transformed pain ratings into higher-pain and lower-pain classes. **e** STN time series signals correspond to recording site. **f** Average power spectral density (PSD) amplitudes for higher- and lower-pain experiences is shown on a log scale for the y-axis, with shaded areas indicating the standard error of the mean (SEM). **g** Observations reordered from lower to higher pain ratings. **h,i** Heatmaps of reordered signal recordings, with z scored PSD values for higher-pain **h** and lower-pain **i** levels within the same frequency range.

## Supplementary Figure S6. Report F

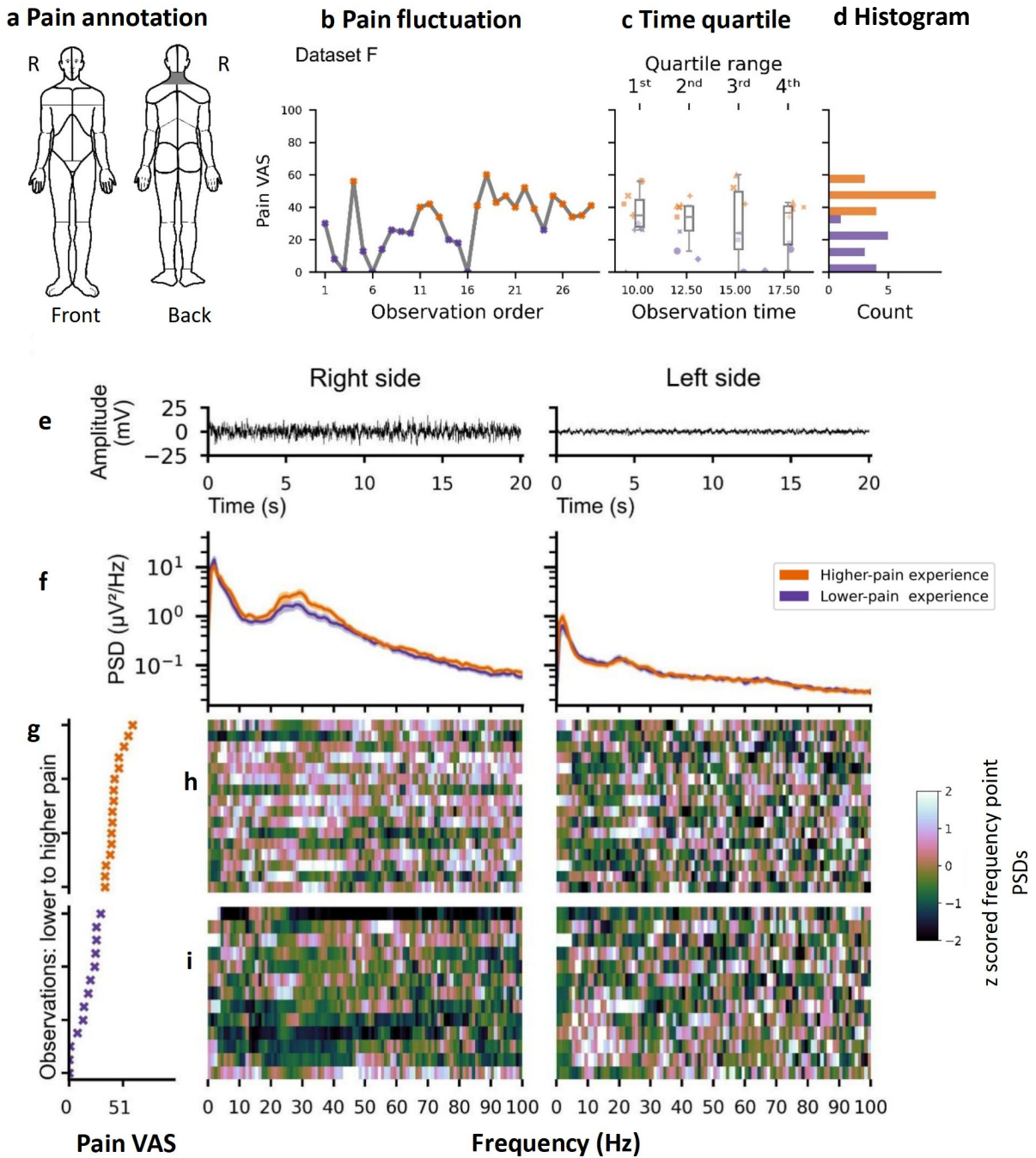

**a** Pain annotation

R  R

Front  Back

**b** Pain fluctuation

Dataset F

**c** Time quartile

Quartile range
1ˢᵗ 2ⁿᵈ 3ʳᵈ 4ᵗʰ

**d** Histogram

**e**

Right side  Left side

**f**

— Higher-pain experience
— Lower-pain experience

**g**

**h**

**i**

z scored frequency point
PSDs

Pain VAS  Frequency (Hz)

**Figure S6.** Individual pain fluctuations in annotated pain report #F. **a** Pain map based on the patient's report, indicating pain in the back of the neck area. The schematic body pain area representation is adapted from Margolis et al.[31] The grey shading represents pain annotations (R: right). **b** Pain fluctuations across the observation sequence. **c** Scatterplot of pain ratings overlaid with daily timestamps for each recording, with same-day recordings marked by identical symbols. Boxplots are shown to represent the central tendency and distribution across four quartiles of the timestamp: median

(middle line), first and third quartiles (box borders), and whiskers extending to 1.5 times the interquartile range. No significant difference was found across the quartiles (Annotated pain report F; $F(28) = 0.10$, $P = 0.96$, one-way ANOVA). **d** Histogram showing pain ratings transformed into higher- and lower-pain classes. **e** STN time series signals correspond to recording site. **f** Average power spectral density (PSD) amplitudes for higher- and lower-pain experiences is shown on a log scale for the y-axis, with shaded areas indicating the standard error of the mean (SEM). **g** Observations reordered from lower to higher pain ratings. **h, i** Heatmaps of reordered signal recordings, with z scored PSD values for higher-pain **h** and lower-pain **i** levels within the same frequency range

## Supplementary Figure S7. Report G

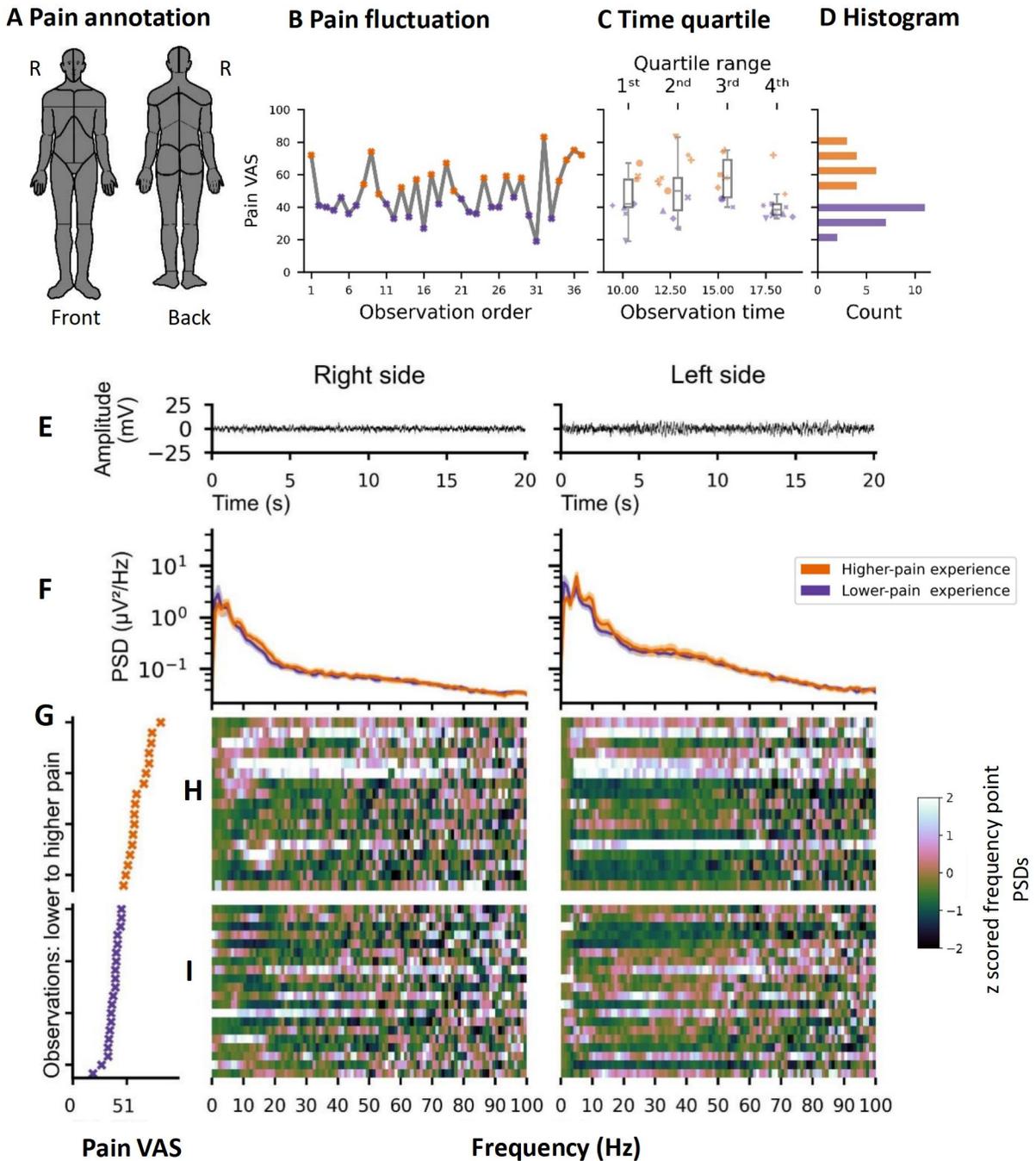

**Figure S7.** Individual pain fluctuations in annotated pain report #G. **a** Pain map refers to the body in general for any pain in any location that the patient felt. The schematic body pain area representation is adapted from Margolis et al.[31] The grey shading represents pain annotations (R: right). **b** Pain fluctuations across the observation sequence. **c** Scatterplot of pain ratings with daily timestamps of data recording, where recordings on the same day are represented by identical symbols. Boxplots display

central tendencies and distributions across four quartiles of the timestamp: median (middle line), first and third quartiles (box borders), and whiskers extending to 1.5 times the interquartile range. No significant differences were observed across quartiles (Annotated pain report G; $F(36) = 2.08$, $P = 0.12$, one-way ANOVA). **d** Histogram showing pain ratings transformed into higher- and lower-pain classes. **e** STN time series signals correspond to recording site. **f** Average power spectral density (PSD) amplitudes for higher- and lower-pain experiences is shown on a log scale for the y-axis, with shaded areas indicating the standard error of the mean (SEM). **g** Observations reordered from lower to higher pain ratings. **h, i** Heatmaps of reordered signal recordings, with z scored PSD values for higher-pain **h** and lower-pain **i** levels within the same frequency range.

## Supplementary Figure S8. Report H

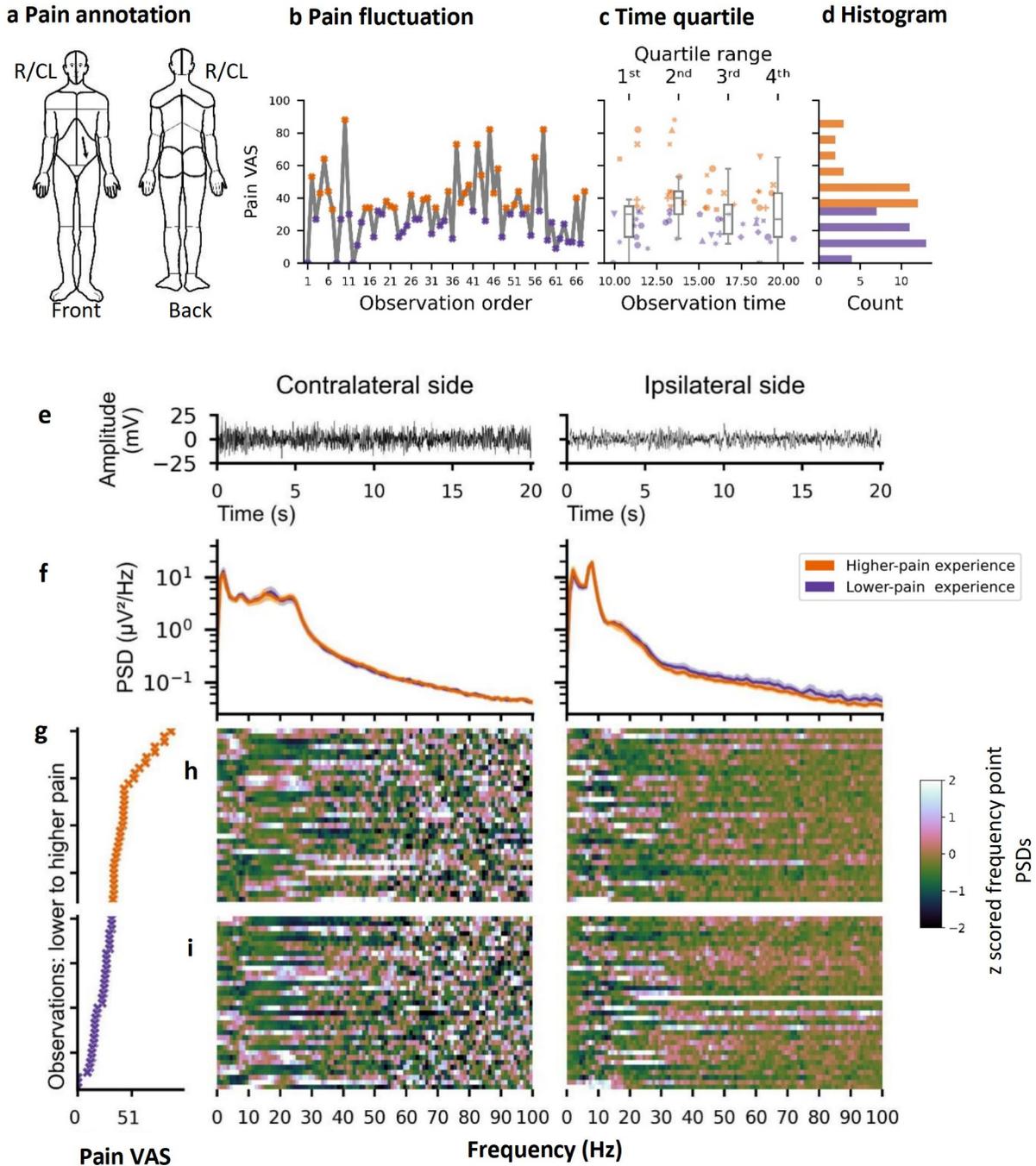

**a** Pain annotation

R/CL

R/CL

Front

Back

**b** Pain fluctuation

**c** Time quartile

**d** Histogram

**e** Contralateral side | Ipsilateral side

**f**

Higher-pain experience
Lower-pain experience

**g**

**h**

**i**

z scored frequency point
PSDs

**Figure S8.** Overview of pain fluctuation report with lateral and specific location and non-PDRP characteristics (annotated pain report #H). **a** Pain map illustrating patient-reported pain in the left lower abdomen, with body pain areas referenced from Margolis et al.[31] Arrows indicate pain locations (R: right; CL: contralateral side). **b** Fluctuation of pain across different observations. **c** Scatterplot of pain ratings by daily recording times; same-day recordings are represented by consistent symbols. Overlaid boxplots display central tendency and distribution across four annotated pain report time quartiles: median (center bar), first and third quartiles (box borders), and whiskers extending to 1.5 times the interquartile

range. One-way ANOVA revealed no significant difference among quartile pain distributions (Annotated pain report H; $F(67) = 2.03$, $P = 0.12$). **d** Histogram classifying pain ratings into higher-pain and lower-pain groups. **e** STN time series signals correspond to recording site. **f** Average power spectral density (PSD) amplitudes for higher and lower pain experiences on a log scale (y-axis); the shaded area represents the standard error of the mean (SEM). **g** Observations reordered from lower to higher pain. **h, i** Heatmaps of z-scored PSDs for reordered observations, showing frequency-specific activity in higher-pain **h** and lower-pain **i** levels.

## Supplementary Figure S9. Pain-level distribution of motor symptoms, mood and

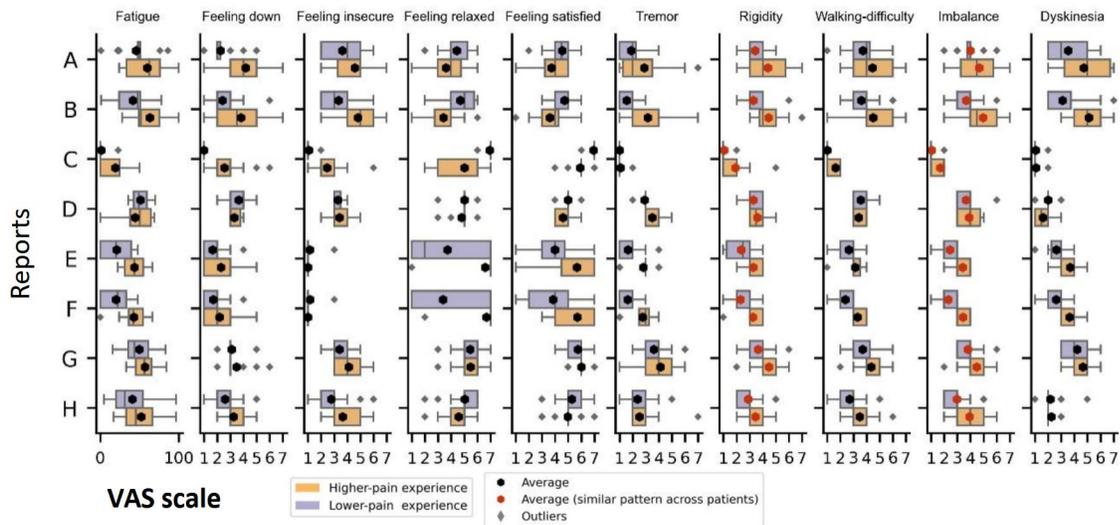

**Figure S9.** Boxplots comparing self-reported rating distributions to corresponding pain severity class labels. Each boxplot represents the distribution of VAS scale ratings, with the median indicated by the middle bar within the box. The lower and upper box borders represent the first and third quartiles, respectively. The top and bottom whiskers extend to 1.5 times the interquartile range, with separate diamonds (◊) marking outlier data points outside the whiskers. Greater pain levels correspond to higher average of *rigidity* and *imbalance* ratings distribution across all annotated pain reports.

**Supplementary Figure S10. Report-specific balanced accuracy performance**

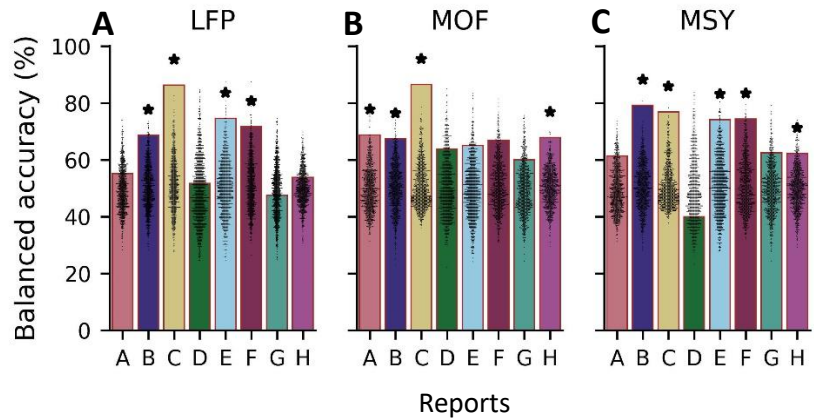

**Figure S10.** Bar plots of random forest classifier performance on the basis of binary pain labels with various features. Each circle (•) represents a single prediction from a shuffled-label annotated pain report, while the bar plot shows the average balanced accuracy across the outer loop of nested cross-validation. A star (★) indicates that a annotated pain report exceeded the one-sided permutation test threshold ($P < 0.05$, n-shuffled = 1000) for the (A) LFP, (B) MOF, and (C) MSY features. *MOF* refers to mood and fatigue, and *MSY* refers to motor symptoms.

# Supplementary Figure S11. Group-level balanced accuracy of motor symptoms, mood, and fatigue.

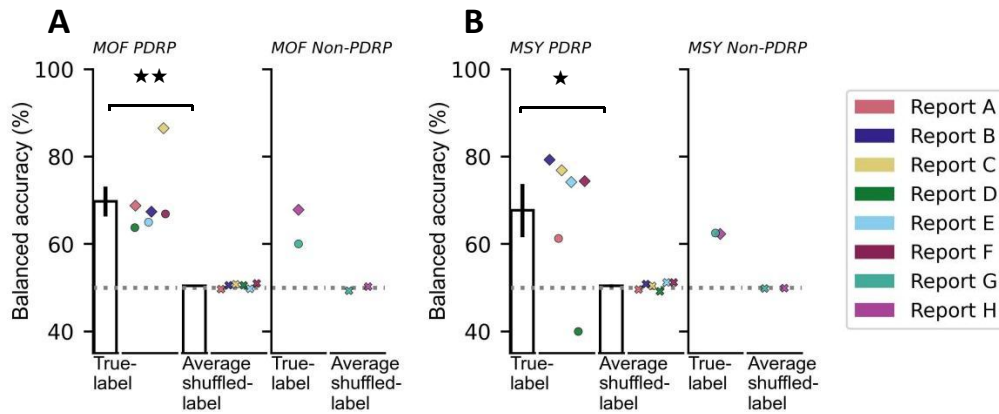

**Figure S11.** Group-level balanced accuracy evaluation for self-reported features. The average balanced accuracy for each annotated pain report is represented by different colours, with features shown in **a** mood and fatigue (MOF) and **b** motor symptoms (MSY). The diamond (◇) denotes annotated pain reports with above-chance performance on the basis of the permutation test, the circle (○) indicates annotated pain reports that do not surpass the permutation test threshold, and the cross (x) represents the average performance from 1000 shuffled-label iterations. The bar plot shows the group-level average of the corresponding PDRP annotated pain reports, with the error bar representing the standard error of the mean (SEM). The star (★★) indicates the results of a one-sided Wilcoxon signed-rank test against the average of the null distribution with P < 0.05/3 (Bonferroni correction), whereas the star (★) represents the results with P < 0.05.

# Supplementary Figure S12. Electrode-specific model performance

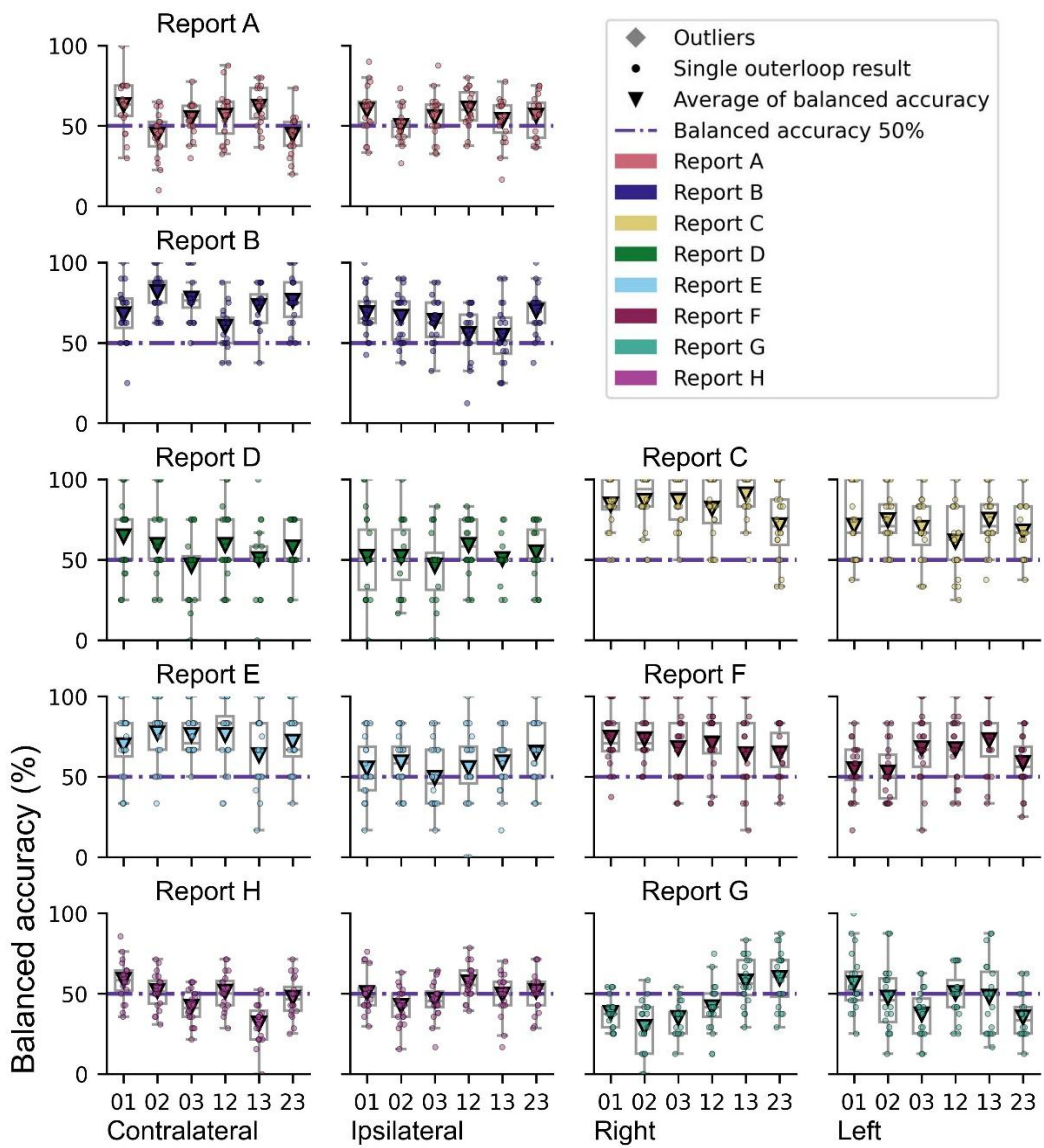

**Figure S12.** Decoding performance of the bipolar electrode model across all annotated pain reports. The boxplots illustrate the distribution of balanced accuracy from nested cross-validation (CV) for the bipolar-electrode models. The middle bar inside the box represents the median, whereas the lower and upper box borders correspond to the first and third quartiles, respectively. The top and bottom whiskers indicate the spread, extending up to 1.5 times the interquartile range.

**Supplementary Figure S13. Group Gini importance of motor symptoms, mood, and fatigue.**

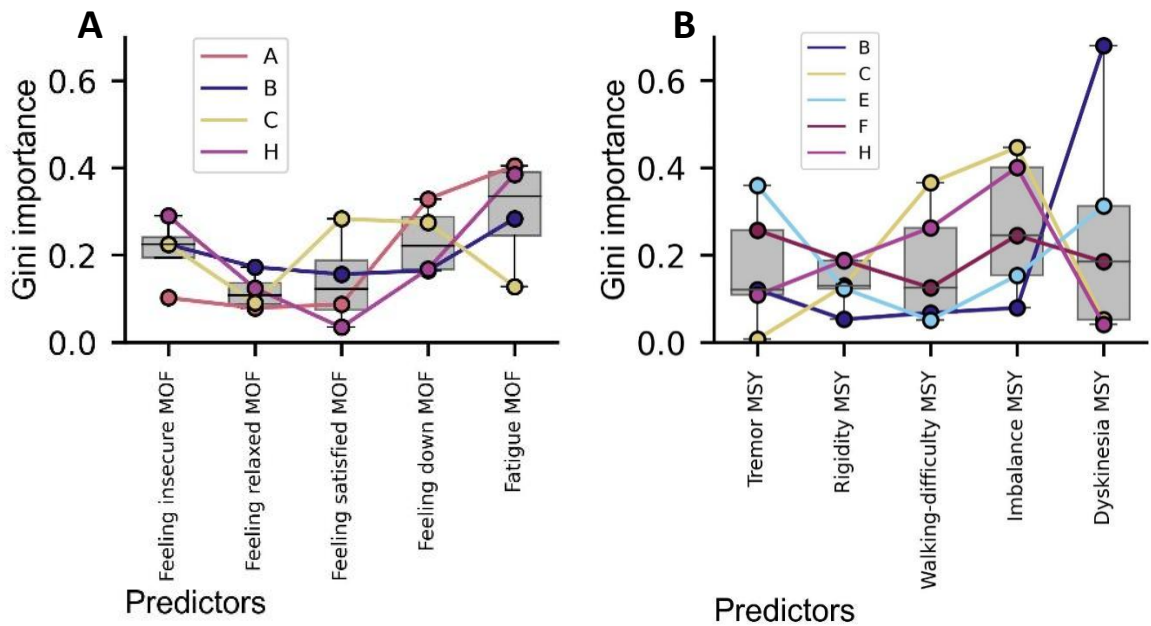

**Figure S13.** Group-level Gini importance of MOF and MSY across all annotated pain reports. (A) Annotated pain reports A, B, C, and H represent annotated pain reports with significant prediction performance using mood and fatigue (MOF) features. (B) Annotated pain reports B, C, E, F, and H are representations of annotated pain reports with significant prediction performance using motor symptoms (MSY) features. The boxplots display the distribution of Gini importance for each annotated pain report: the median is represented by a middle bar in the box, with the lower and upper borders denoting the first and third quartiles, respectively. The top and bottom whiskers extend to 1.5 times the interquartile range, and separate points outside the whiskers represent outlier data points. MOF: Mood and fatigue; MSY: Motor symptom

# Supplementary Figure S14

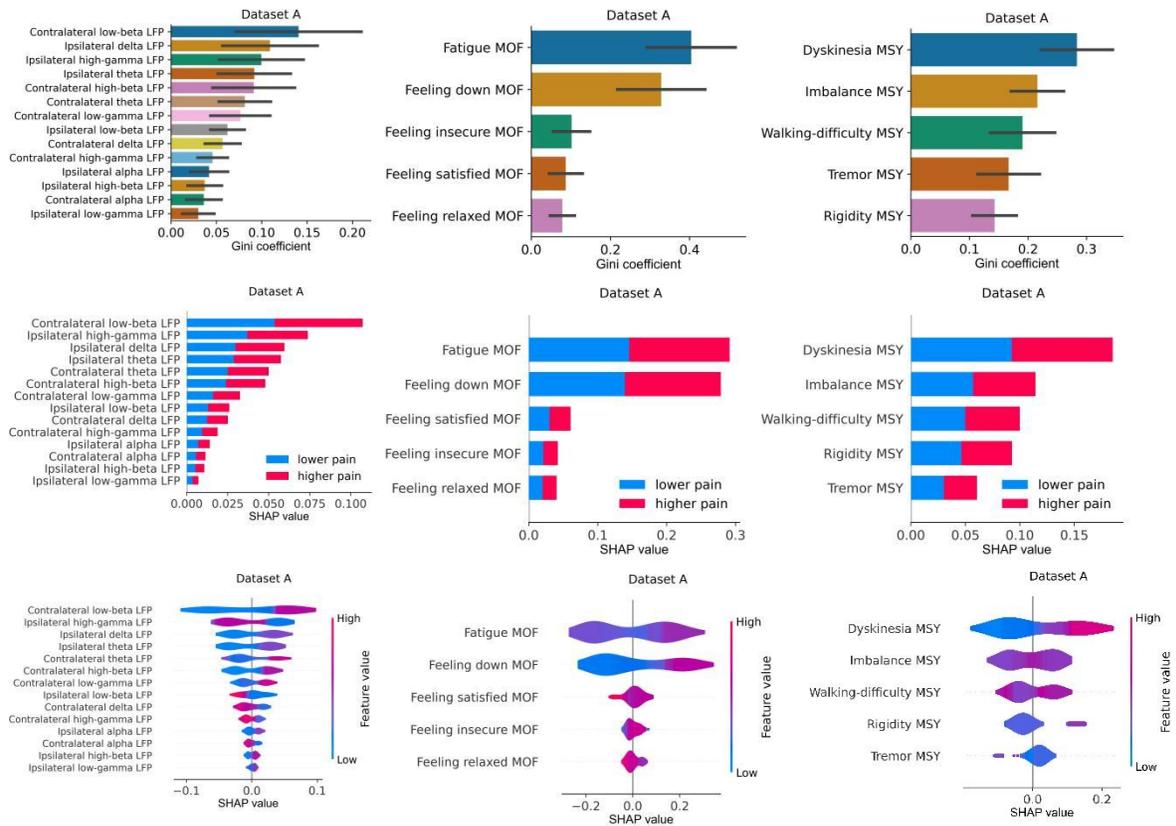

**Figure S14. Annotated pain report A.** Gini importance and SHAP as indicators of feature influence. The figure presents model explainability via the Gini importance and SHAP values for feature influence. Upper row: Bar plots showing feature importance on the basis of the average Gini importance across the outer loop of nested cross-validation (CV), with error bars representing standard deviation (+SD). Middle row: Bar plots of feature importance are ordered in descending order on the basis of retrained SHAP values across all annotated pain report observations using the best hyperparameters for random forests. Bottom row: Violin plots of feature importance in descending order from retrained SHAP values, calculated for all annotated pain report observations with the best random forest hyperparameters. Columns: The left column represents the LFP features, the middle column shows the MOF features, and the right column displays the MSY features. MOF: Mood and fatigue; MSY: Motor symptoms

# Supplementary Figure S15

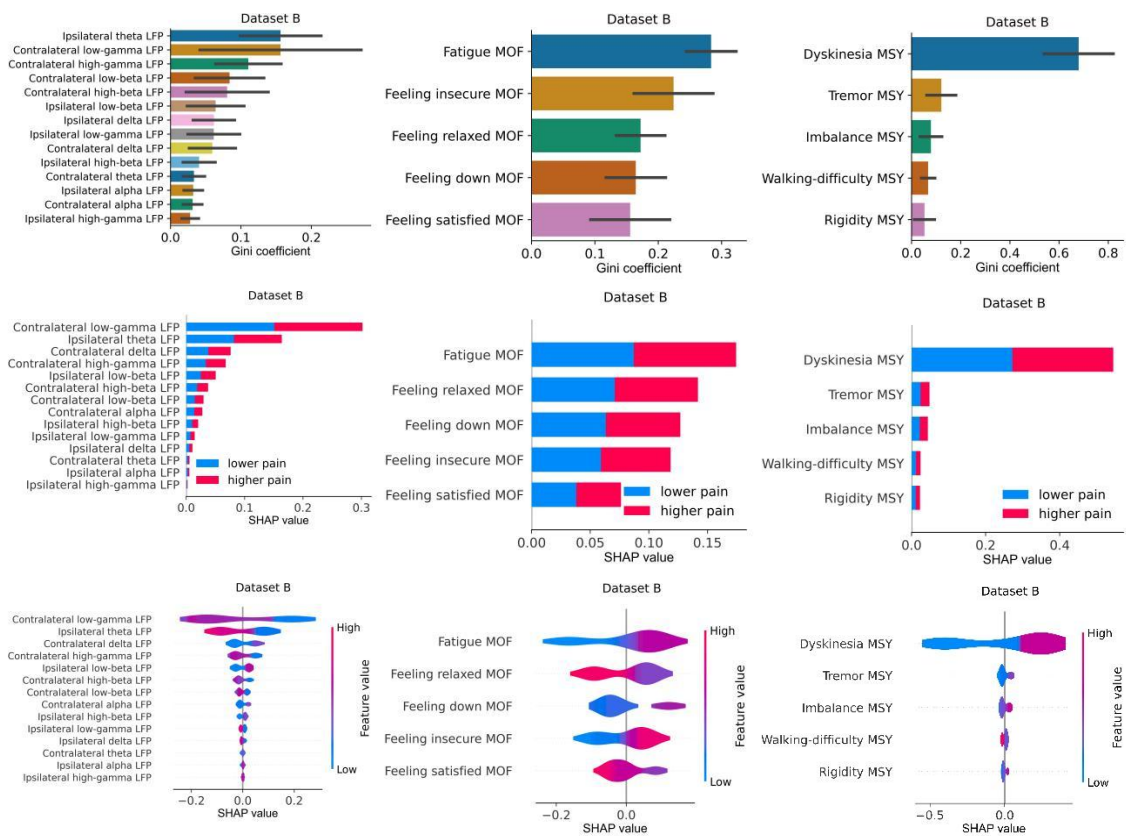

**Figure S15. Annotated pain report B.** Gini importance and SHAP as indicators of feature influence. The figure presents model explainability via the Gini importance and SHAP values for feature influence. Upper row: Bar plots showing feature importance on the basis of the average Gini importance across the outer loop of nested cross-validation (CV), with error bars representing ±SD. Middle row: Bar plots of feature importance are ordered in descending order on the basis of retrained SHAP values across all annotated pain report observations using the best hyperparameters for random forests. Bottom row: Violin plots of feature importance in descending order from retrained SHAP values, calculated for all annotated pain report observations with the best random forest hyperparameters. Columns: The left column represents LFP features, the middle column shows MOF features, and the right column displays MSY features. MOF: Mood and fatigue; MSY: Motor symptoms

## Supplementary Figure S16

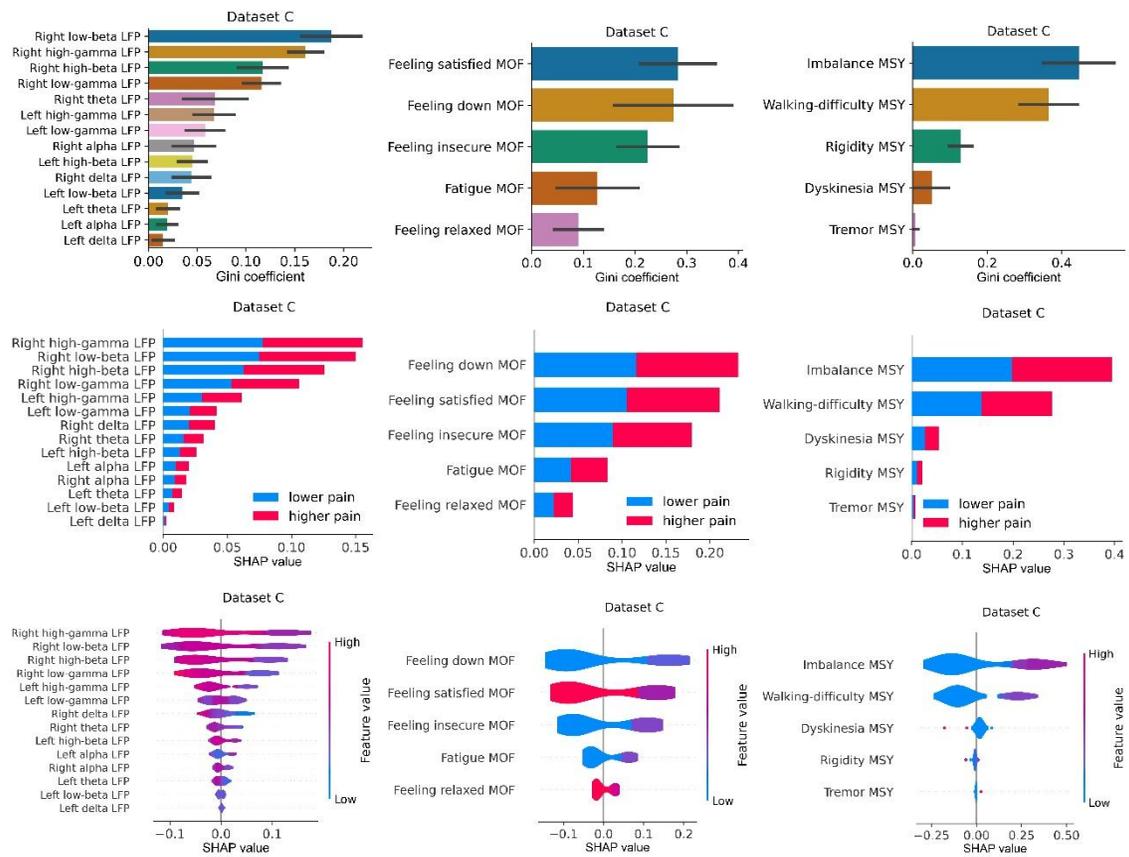

**Figure S16. Annotated pain report C.** Gini importance and SHAP as indicators of feature influence. The figure presents model explainability via the Gini importance and SHAP values for feature influence. Upper row: Bar plots showing feature importance on the basis of the average Gini importance across the outer loop of nested cross-validation (CV), with error bars representing ±SD. Middle row: Bar plots of feature importance are ordered in descending order on the basis of retrained SHAP values across all annotated pain report observations using the best hyperparameters for random forests. Bottom row: Violin plots of feature importance in descending order from retrained SHAP values, calculated for all annotated pain report observations with the best random forest hyperparameters. Columns: The left column represents LFP features, the middle column shows MOF features, and the right column displays MSY features. MOF: Mood and fatigue. MSY: Motor symptoms.

**Supplementary Figure S17**

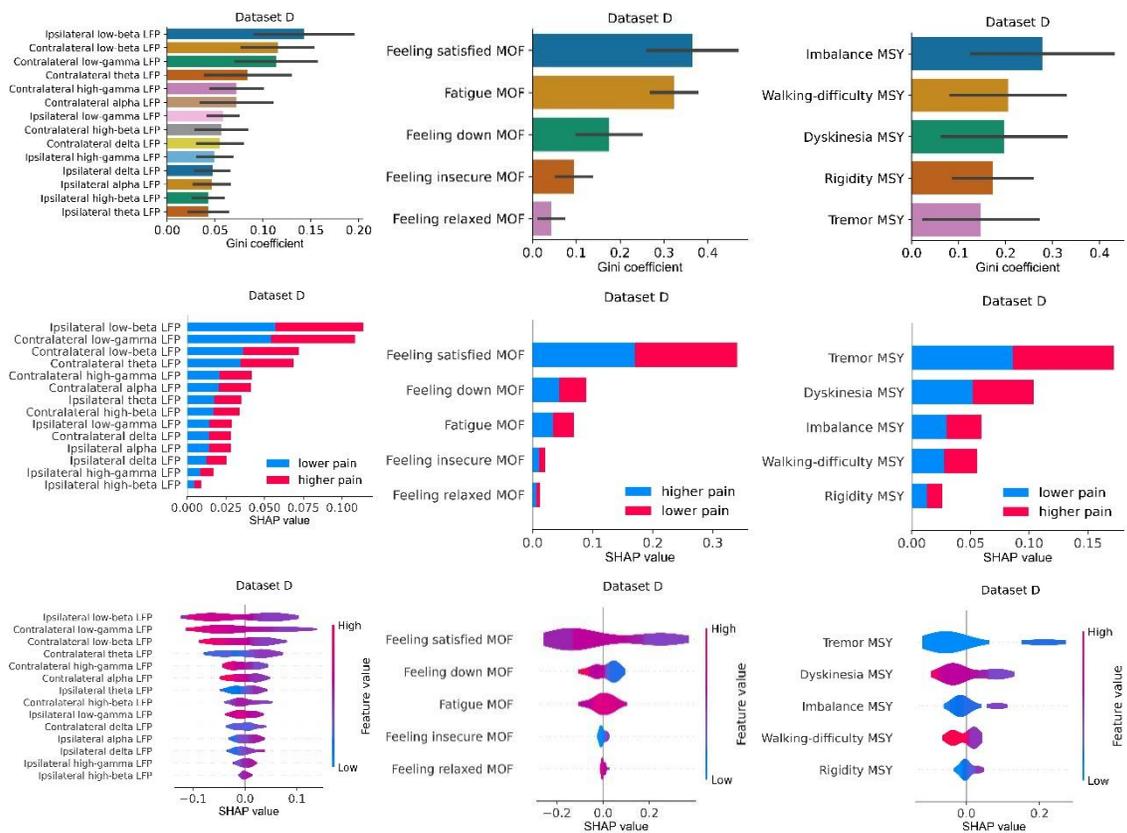

**Figure S17. Annotated pain report D.** Gini importance and SHAP as model explainability for feature influence. The figure presents model explainability via the Gini importance and SHAP values for feature influence. Upper row: Bar plots showing feature importance on the basis of the average Gini importance across the outer loop of nested cross-validation (CV), with error bars representing ±SD. Middle row: Bar plots of feature importance are ordered in descending order on the basis of retrained SHAP values across all annotated pain report observations using the best hyperparameters for random forests. Bottom row: Violin plots of feature importance in descending order from retrained SHAP values, calculated for all annotated pain report observations with the best random forest hyperparameters. Columns: The left column represents LFP features, the middle column shows MOF features, and the right column displays MSY features. MOF: Mood and fatigue; MSY: Motor symptoms.

## Supplementary Figure S18

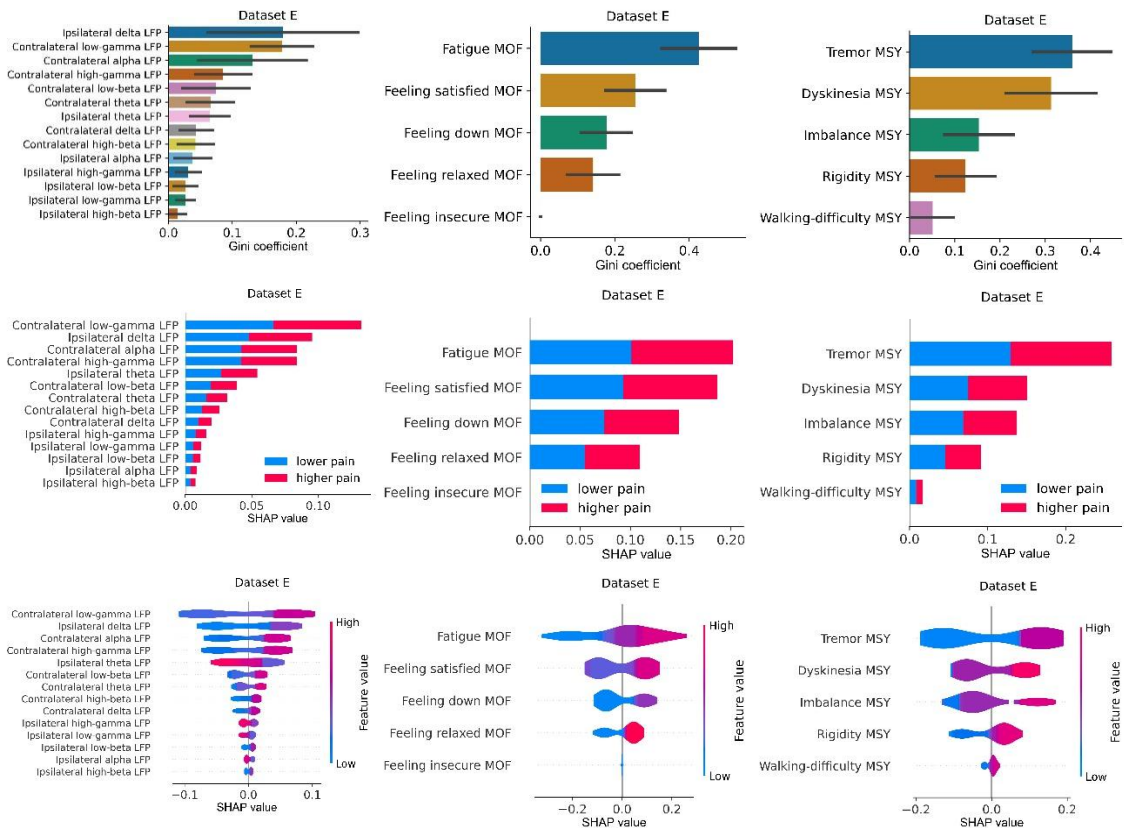

**Figure S18. Annotated pain report E.** Gini importance and SHAP as model explainability for feature influence. The figure presents model explainability via the Gini importance and SHAP values for feature influence. Upper row: Bar plots showing feature importance on the basis of the average Gini importance across the outer loop of nested cross-validation (CV), with error bars representing ±SD. Middle row: Bar plots of feature importance are ordered in descending order on the basis of retrained SHAP values across all annotated pain report observations using the best hyperparameters for random forests. Bottom row: Violin plots of feature importance in descending order from retrained SHAP values, calculated for all annotated pain report observations with the best random forest hyperparameters. Columns: The left column represents LFP features, the middle column shows MOF features, and the right column displays MSY features. MOF: Mood and fatigue; MSY: Motor symptoms.

**Supplementary Figure S19**

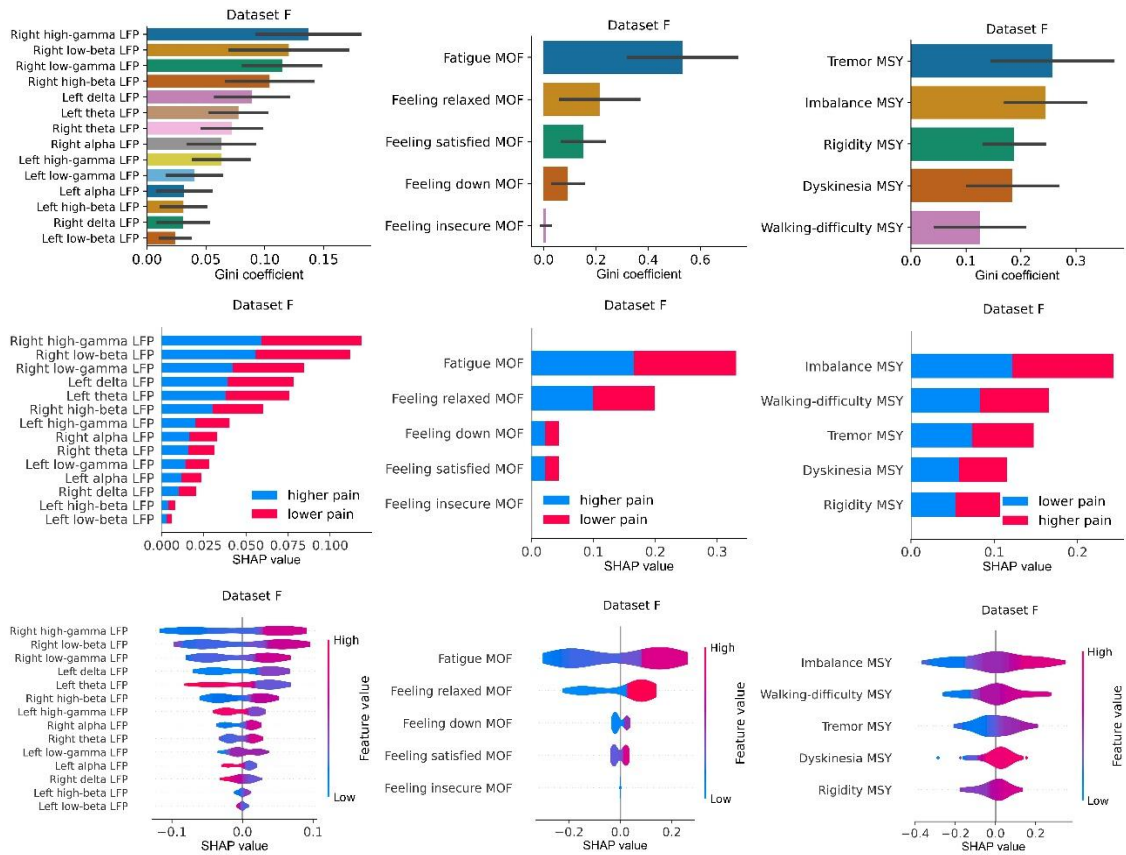

**Figure S19. Annotated pain report F.** Gini importance and SHAP as model explainability for feature influence. The figure presents model explainability via the Gini importance and SHAP values for feature influence. Upper row: Bar plots showing feature importance on the basis of the average Gini importance across the outer loop of nested cross-validation (CV), with error bars representing ±SD. Middle row: Bar plots of feature importance are ordered in descending order on the basis of retrained SHAP values across all annotated pain report observations using the best hyperparameters for random forests. Bottom row: Violin plots of feature importance in descending order from retrained SHAP values, calculated for all annotated pain report observations with the best random forest hyperparameters. Columns: The left column represents LFP features, the middle column shows MOF features, and the right column displays MSY features. MOF: Mood and fatigue; MSY: Motor symptoms.

# Supplementary Figure S20

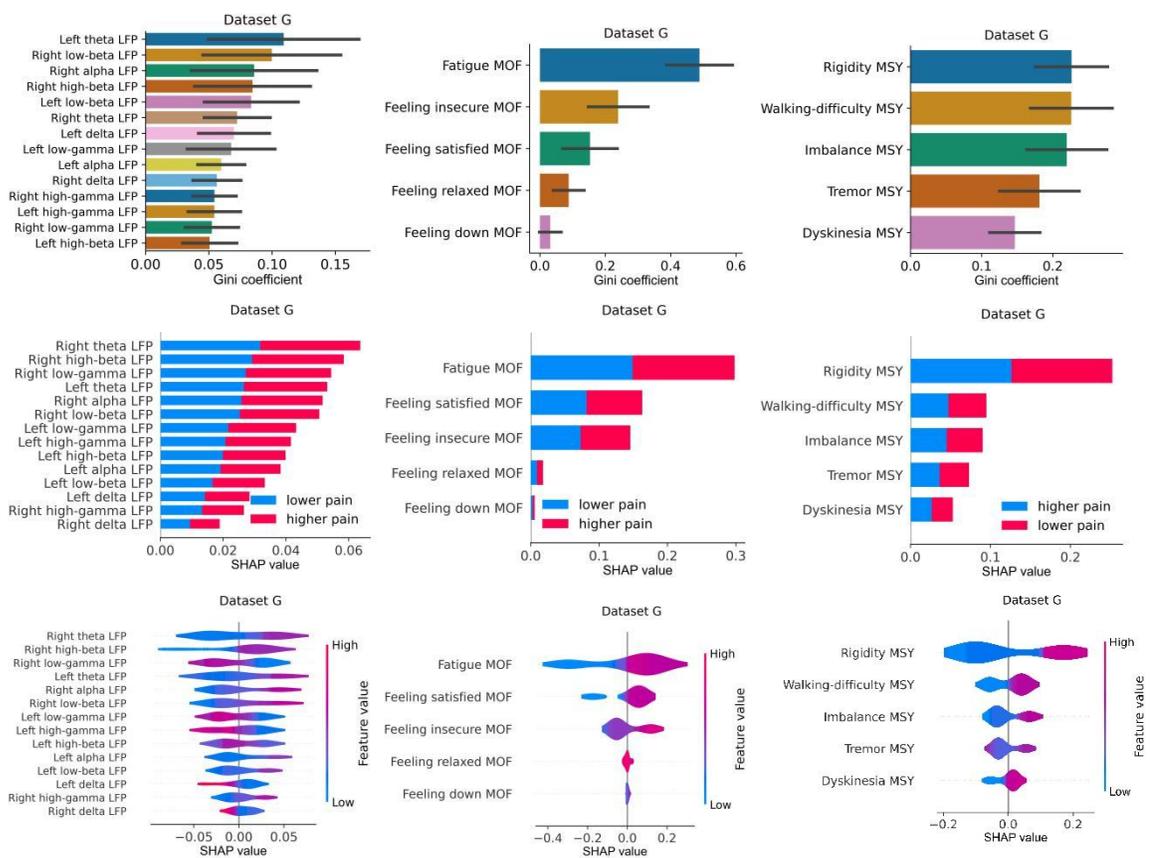

**Figure S20. Annotated pain report G.** Gini importance and SHAP as model explainability on feature influence. The figure presents model explainability via the Gini importance and SHAP values for feature influence. Upper row: Bar plots showing feature importance on the basis of the average Gini importance across the outer loop of nested cross-validation (CV), with error bars representing +SD. Middle row: Bar plots of feature importance are ordered in descending order on the basis of retrained SHAP values across all annotated pain report observations using the best hyperparameters for random forests. Bottom row: Violin plots of feature importance in descending order from retrained SHAP values, calculated for all annotated pain report observations with the best random forest hyperparameters. Columns: The left column represents LFP features, the middle column shows MOF features, and the right column displays MSY features. MOF: Mood and fatigue; MSY: Motor symptoms.

# Supplementary Figure S21

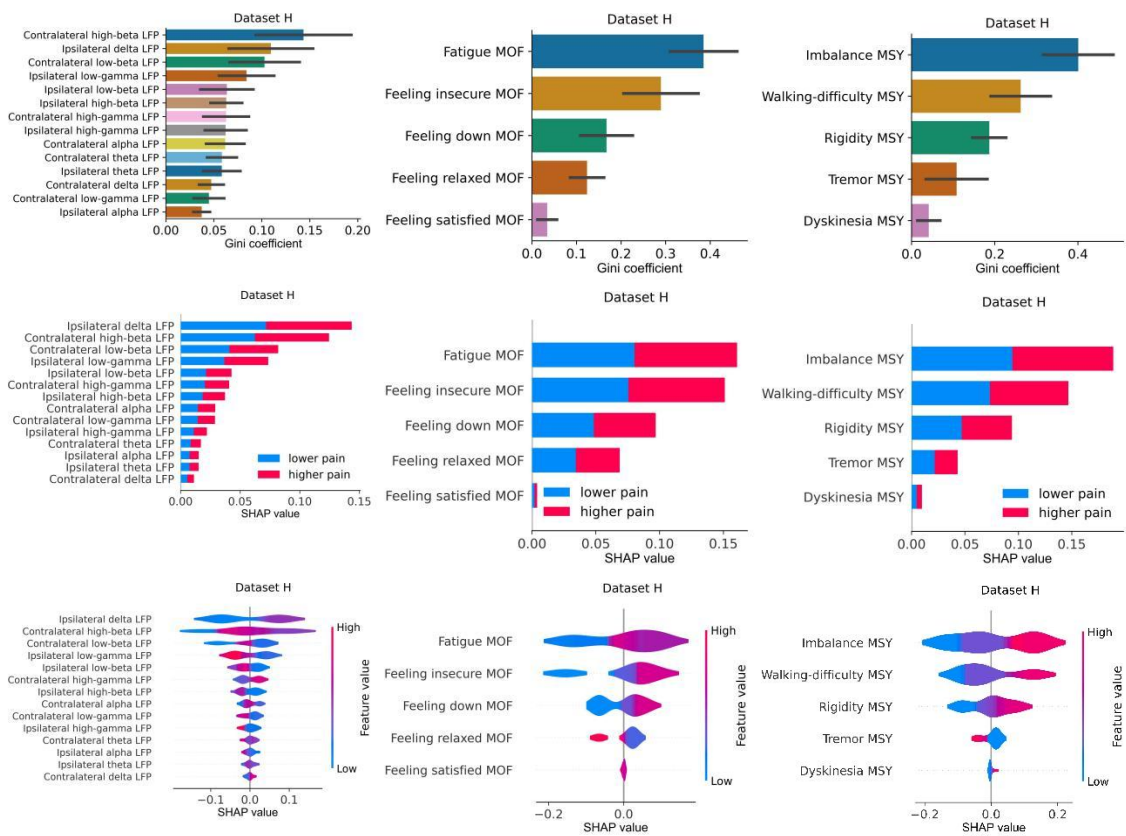

**Figure S21. Annotated pain report H.** Gini importance and SHAP as indicators of feature influence. The figure presents model explainability via the Gini importance and SHAP values for feature influence. Upper row: Bar plots showing feature importance on the basis of the average Gini importance across the outer loop of nested cross-validation (CV), with error bars representing ±SD. Middle row: Bar plots of feature importance are ordered in descending order on the basis of retrained SHAP values across all annotated pain report observations using the best hyperparameters for random forests. Bottom row: Violin plots of feature importance in descending order from retrained SHAP values, calculated for all annotated pain report observations with the best random forest hyperparameters. Columns: The left column represents LFP features, the middle column shows MOF features, and the right column displays MSY features. MOF: Mood and fatigue; MSY: Motor symptoms.

**Supplementary Figure S22. Schematic reports collection**

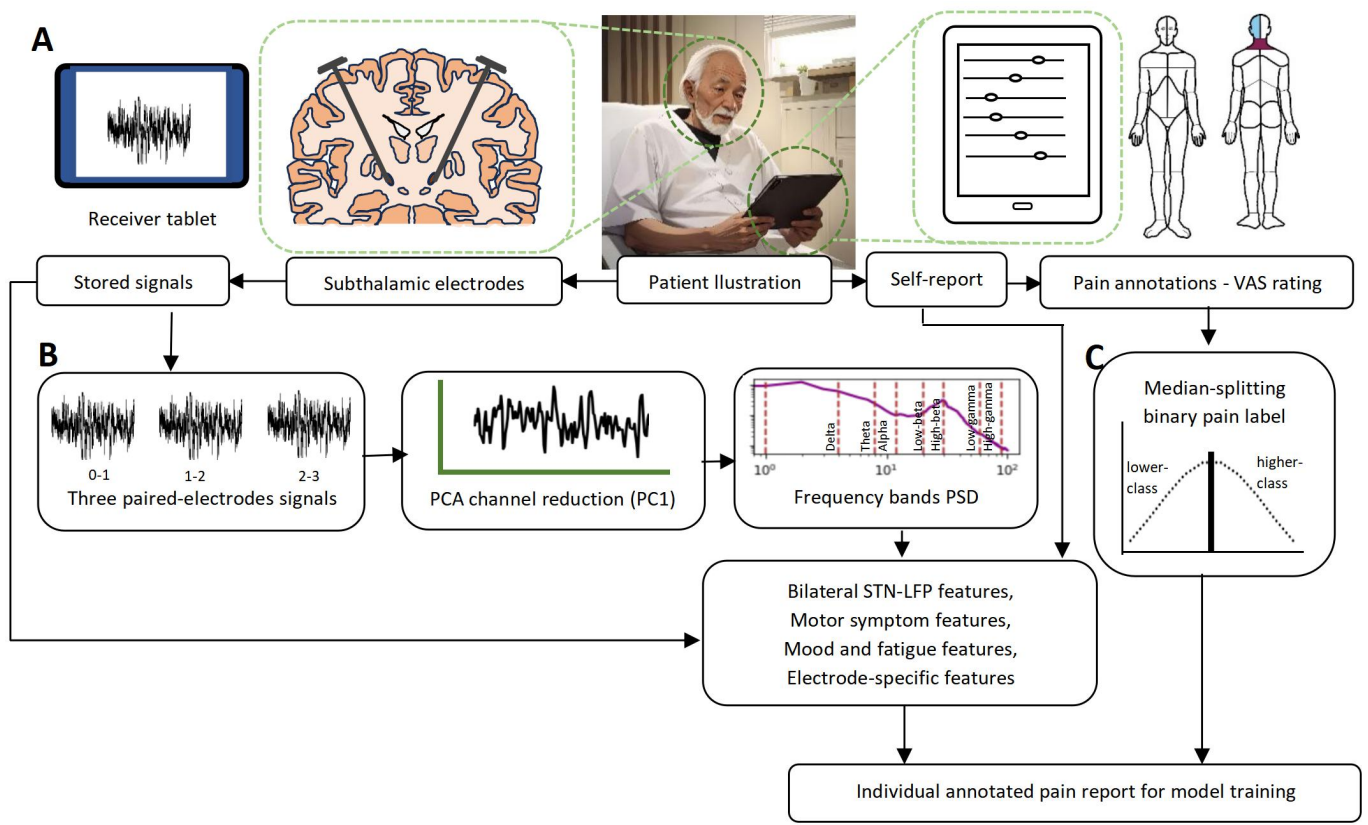

**Figure S22.** Schematic representation of intraindividual multimodal annotated pain report collection for binary pain fluctuation classification. **a** Pain ratings were collected via a tablet PC (note: patient image derived from text-to-image artificial intelligence tool via www.fotor.com). Implanted electrodes targeting the subthalamic nucleus (STN) recorded neural signals, which were transmitted wirelessly via the Percept™ PC deep brain stimulation (DBS) system (Medtronic, Inc.). This system was activated through the BrainSense™ Survey. **b** Signals from each STN hemisphere, comprising three electrode pairs (0-1, 1-2, 2-3), were captured via the BrainSense™ Survey. A single time series was retained via the first principal component (PC1) from principal component analysis (PCA). The resulting time series were filtered and transformed to power spectral density (PSD), allowing the calculation of average PSD amplitudes across the delta, theta, alpha, low-beta, high-beta, low-gamma, and high-gamma frequency bands. The BrainSense™ Survey also provides PSD data for each electrode pair across frequency bands. **c** Features were compiled for model training. These features include STN local field potentials (STN-LFPs), motor symptom ratings (MSY), and mood and fatigue assessments (MOF). Bipolar electrode frequency bands were used in an additional model for evaluation. **d** Pain ratings associated with each annotation were converted to binary labels (higher vs. lower pain) on the basis of the median rating. Paired feature data and pain ratings from each annotation formed the training annotated pain report for the pain fluctuation classification models.

**SUPPLEMENTARY TABLES**

**Supplementary Table S1.** PD-pain association type evaluation

| No | Assessment | Annotated pain report | | | | | | | |
|---|---|---|---|---|---|---|---|---|---|
| | | A | B | C | D | E | F | G | H |
| 1 | Pain annotation with specific location | Yes | Yes | Yes | Yes | Yes | Yes | No | Yes |
| 2 | Apparent cause of pain other than PD | No | No | No | No | No | No | | Yes |
| 3 | Temporal relationship with the course of disease | Yes | Yes | Yes | No | Yes | Yes | | |
| 4 | Impact of motor fluctuation (rigidity, tremors, slowness of movements) | No | No | No | Yes | No | No | | |
| 5 | Impact of motor fluctuation (excessive, abnormal movements [choreatic dyskinesia]) | No | No | No | No | No | No | | |
| 6 | Impact of Parkinsonian treatment | Yes | Yes | Yes | No | Yes. | Yes | | |
| **Group** | | PDRP | PDRP | PDRP | PDRP | PDRP | PDRP | Non-PDRP | Non-PDRP |

**Supplementary Table S2.** Self-report and rating of the symptoms

| Domain | Description | Score | Description | Role |
|---|---|---|---|---|
| **Mood** | I feel insecure | 1-7 | 1= not at all, 4= moderate, 7= the most | feature |
| **Mood** | I feel relaxed | 1-7 | 1= not at all, 4= moderate, 7= the most | feature |
| **Mood** | I feel satisfied | 1-7 | 1= not at all, 4= moderate, 7= the most | feature |
| **Mood** | I feel down | 1-7 | 1= not at all, 4= moderate, 7= the most | feature |
| **PD motor symptom** | I experience tremor | 1-7 | 1= not at all, 4= moderate, 7= the most | feature |
| **PD motor symptom** | I experience rigidity | 1-7 | 1= not at all, 4= moderate, 7= the most | feature |
| **PD motor symptom** | Walking is difficult | 1-7 | 1= not at all, 4= moderate, 7= the most | feature |
| **PD motor symptom** | I experience balance problems | 1-7 | 1= not at all, 4= moderate, 7= the most | feature |
| **PD motor symptom** | I experience dyskinesia | 1-7 | 1= not at all, 4= moderate, 7= the most | feature |
| **Fatigue** | I feel fatigue | 0-100 | 0= not at all, 50= moderate, 100= the most | feature |
| **Pain severity** | I feel pain (prespecified location/with annotation) | 0-100 | 0= not at all, 50= moderate, 100= the most | Target label |

# Reference

1.  Feldmann, L. K. *et al.* Toward therapeutic electrophysiology: beta-band suppression as a biomarker in chronic local field potential recordings. *Npj Park. Dis.* **8**, 44 (2022).

2.  Mamun, K. A. *et al.* Movement decoding using neural synchronization and inter-hemispheric connectivity from deep brain local field potentials. *J. Neural Eng.* **12**, 056011 (2015).

3.  Thompson, J. A. *et al.* Sleep patterns in Parkinson's disease: direct recordings from the subthalamic nucleus. *J. Neurol. Neurosurg. Psychiatry* **89**, 95–104 (2018).

4.  Verma, A. K. *et al.* Parkinsonian daytime sleep-wake classification using deep brain stimulation lead recordings. *Neurobiol. Dis.* **176**, 105963 (2023).

5.  Sand, D. *et al.* Machine learning-based personalized subthalamic biomarkers predict ON-OFF levodopa states in Parkinson patients. *J. Neural Eng.* **18**, 046058 (2021).

6.  McNamara, P., Stavitsky, K., Harris, E., Szent-Imrey, O. & Durso, R. Mood, side of motor symptom onset and pain complaints in Parkinson's disease. *Int. J. Geriatr. Psychiatry* **25**, 519–524 (2010).

7.  Berardelli, A. *et al.* Pathophysiology of pain and fatigue in Parkinson's disease. *Parkinsonism Relat. Disord.* **18**, S226–S228 (2012).

8.  Hagell, P. & Brundin, L. Towards an understanding of fatigue in Parkinson disease. *J. Neurol. Neurosurg. Psychiatry* **80**, 489–492 (2009).

9.  Cattaneo, C. *et al.* Long-Term Effects of Safinamide on Mood Fluctuations in Parkinson's Disease. *J. Park. Dis.* **7**, 629–634 (2017).

10. Pauletti, C. *et al.* Fatigue in fluctuating Parkinson's disease patients: possible impact of safinamide. *J. Neural Transm.* **130**, 915–923 (2023).

11. Jasti, S., Dudley, W. N. & Goldwater, E. SAS Macros for Testing Statistical Mediation in Data With Binary Mediators or Outcomes. *Nurs. Res.* **57**, 118–122 (2008).

12. Jacobson, C. J. *et al.* Disclosure and Self-Report of Emotional, Social, and Physical Health in Children and Adolescents With Chronic Pain—A Qualitative Study of PROMIS Pediatric Measures. *J. Pediatr. Psychol.* **38**, 82–93 (2013).

13. Villemure, C. & Bushnell, M. C. Mood Influences Supraspinal Pain Processing Separately from Attention. *J. Neurosci.* **29**, 705–715 (2009).

14. Rana, A. Q. *et al.* Association of restless legs syndrome, pain, and mood disorders in parkinson's disease. *Int. J. Neurosci.* **126**, 116–120 (2016).

15. Lim, S.-Y. *et al.* Do dyskinesia and pain share common pathophysiological mechanisms in Parkinson's disease? *Mov. Disord.* **23**, 1689–1695 (2008).

16. Ford, B. Pain in Parkinson's disease. *Mov. Disord.* **25**, S98–S103 (2010).

17. Tinazzi, M. *et al.* Pain and motor complications in Parkinson's disease. *J. Neurol. Neurosurg. Psychiatry* **77**, 822–825 (2006).

18. Yin, Z. *et al.* Local field potentials in Parkinson's disease: A frequency-based review. *Neurobiol. Dis.* **155**, 105372 (2021).

19. Valkovic, P. *et al.* Pain in Parkinson´s Disease: A Cross-Sectional Study of Its Prevalence, Types, and Relationship to Depression and Quality of Life. *PLOS ONE* **10**, e0136541 (2015).

20. Broetz, D., Eichner, M., Gasser, T., Weller, M. & Steinbach, J. P. Radicular and nonradicular back pain in Parkinson's disease: A controlled study. *Mov. Disord.* **22**, 853–856 (2007).

21. Zambito Marsala, S. *et al.* Spontaneous pain, pain threshold, and pain tolerance in Parkinson's disease. *J. Neurol.* **258**, 627–633 (2011).

22. Magariños-Ascone, C. M., Figueiras-Mendez, R., Riva-Meana, C. & Córdoba-Fernández, A. Subthalamic neuron activity related to tremor and movement in Parkinson's disease. *Eur. J. Neurosci.* **12**, 2597–2607 (2000).

23. Narabayashi, Y. & Oshima, T. Analysis of parkinsonian tremor and rigidity relying on peripheral and central intensifying mechanisms through thalamic activities. *Neurol. Clin. Neurosci.* **2**, 28–37 (2014).

24. Lee, M. A., Walker, R. W., Hildreth, T. J. & Prentice, W. M. A Survey of Pain in Idiopathic Parkinson's Disease. *J. Pain Symptom Manage.* **32**, 462–469 (2006).

25. Mylius, V., Möller, J. C., Bohlhalter, S., Ciampi de Andrade, D. & Perez Lloret, S. Diagnosis and Management of Pain in Parkinson's Disease: A New Approach. *Drugs Aging* **38**, 559–577 (2021).

26. Nègre-Pagès, L., Regragui, W., Bouhassira, D., Grandjean, H. & Rascol, O. Chronic pain in Parkinson's disease: The cross-sectional French DoPaMiP survey. *Mov. Disord.* **23**, 1361–1369 (2008).

27. Mylius, V. *et al.* Pain in Parkinson's Disease: Current Concepts and a New Diagnostic Algorithm. *Mov. Disord. Clin. Pract.* **2**, 357–364 (2015).

28. Bruno, A. E. & Sethares, K. A. Fatigue in Parkinson Disease: An Integrative Review. *J. Neurosci. Nurs.* **47**, 146–153 (2015).

29. Lin, I. *et al.* Triggers and alleviating factors for fatigue in Parkinson's disease. *PLOS ONE* **16**, e0245285 (2021).

30. Zarotti, N. *et al.* Psychological interventions for people with Parkinson's disease in the early 2020s: Where do we stand? *Psychol. Psychother. Theory Res. Pract.* **94**, 760–797 (2021).

31. Margolis, R. B., Tait, R. C. & Krause, S. J. A rating system for use with patient pain drawings: *Pain* **24**, 57–65 (1986).